\documentclass[aps,prl,twocolumn,groupedaddress,showpacs,floatfix]{revtex4}
\usepackage{dcolumn}
\usepackage{graphicx}
\usepackage{amsmath}
\usepackage{amssymb}
\usepackage{color}
\usepackage{bm}       
\usepackage{flafter}
\usepackage{epsfig}
\usepackage{epstopdf}
\usepackage{float}
\usepackage{multirow}

\begin{document}


\title{Magnetic structure, magnetoelastic coupling, and thermal properties of EuCrO$_3$ nano-powders}

\author{M. Taheri}
\affiliation{ Department of Physics, Brock University, St Catharines, ON, L2S 3A1, Canada}
\author{F. S. Razavi}
\affiliation{ Department of Physics, Brock University, St Catharines, ON, L2S 3A1, Canada}
\author{Z. Yamani}
\affiliation{ Canadian Neutron Beam Centre, Chalk River Laboratories, Chalk River, ON, K0J 1J0, Canada}
\author{R. Flacau}
\affiliation{ Canadian Neutron Beam Centre, Chalk River Laboratories, Chalk River, ON, K0J 1J0, Canada}
\author{P. G. Reuvekamp}
\affiliation{ Max Planck Institute for Solid State Research, Stuttgart, Germany}
\author{A. Schulz}
\affiliation{ Max Planck Institute for Solid State Research, Stuttgart, Germany}
\author{R. K. Kremer}
\affiliation{ Max Planck Institute for Solid State Research, Stuttgart, Germany}

\begin{abstract}
We carried out  detailed studies of the magnetic structure, magnetoelastic coupling, and thermal properties of EuCrO$_3$ nano-powders from room temperature to liquid helium temperature. Our neutron powder diffraction and X-ray powder diffraction  measurements provide precise atomic positions of all atoms in the cell, especially for the light oxygen atoms. The low-temperature  neutron powder diffraction data revealed extra Bragg peaks of magnetic origin which can be attributed to a $G_x$ antiferromagnetic structure with an ordered moment of $\sim$ 2.4 $\mu_{\rm B}$ consistent with the $3d^3$ electronic configuration of the Cr$^{3+}$ cations. Apart from previously reported  antiferromagnetic and ferromagnetic transitions in EuCrO$_3$ at low temperatures,  we also observed an anomaly at about 100 K. This anomaly was observed in temperature dependence of sample's, lattice parameters, thermal expansion, Raman spectroscopy, permittivity and conductance measurements. This anomaly is attributed to the magnetoelastic distortion in the EuCrO$_3$ crystal.
\end{abstract}


\maketitle

\section{Introduction}

Since the early investigation of the antiferromagnetic   structure  of LaCrO$_3$ ($T_{\rm N} \sim$ 320 K) by Koehler and Wollan\cite{Koehler1957}  rare earth orthochromites with the chemical composition RCrO$_3$,  (R = rare earth element)  have experienced particular attention. Interest was especially triggered by their complex magnetic ordering often with spin canting resulting from the interaction of the magnetic moments of the Cr$^{3+}$ and the R$^{3+}$ cations which can be tuned by external parameters like temperature, magnetic field or pressure.\cite{Hornreich1978} As such the rare earth orthochromites were perceived as  systems with great potential as magnetoelectric multiferroics (ME/MF) materials. In fact, the rare earth orthochromites have  been demonstrated to exhibit  ferroelectric polarization either induced by an external magnetic field (ME) or a spontaneous polarization as a consequence of internal magnetic fields induced by long-range magnetic ordering (MF). \cite{Sahu2007,Rajeswaran2012,Raveau2014,Saha2014,Meher2014,McDannald2015}

The potential of the RCrO$_3$ was especially seen in their high magnetic ordering temperatures compared to what was observed in the orthomanganites series, RMnO$_3$, (e.g. $T_{\rm N} \sim$ 27 K in TbMnO$_3$) which brought about some of the most prominent MF systems. In comparison to the rare earth orthoferrites, RFeO$_3$,  which exhibit even higher magnetic ordering temperatures, the orthochromites,  however, are potentially more interesting
since the coupling between the rare earth and the transition metal subsystem is larger in magnitude allowing to take advantage of the partially sizeable magnetic moments of the R$^{3+}$ cations. Due to the interaction of the rare earth and 3$d$ moments, RCrO$_3$ systems exhibit a variety of complex magnetic structures, often with the presence of a weak ferromagnetic component and several spontaneous spin reorientations.\cite{Rajeswaran2012} These
were ascribed to the absence of fourth-order crystal field splitting terms of the $S$ =3/2 spin multiplet of the Cr$^{3+}$ cations with an 3$d^3$ electronic configuration.\cite{Hornreich1978}

\begin{figure}[htc]
	\centering
   \includegraphics[height=6cm]{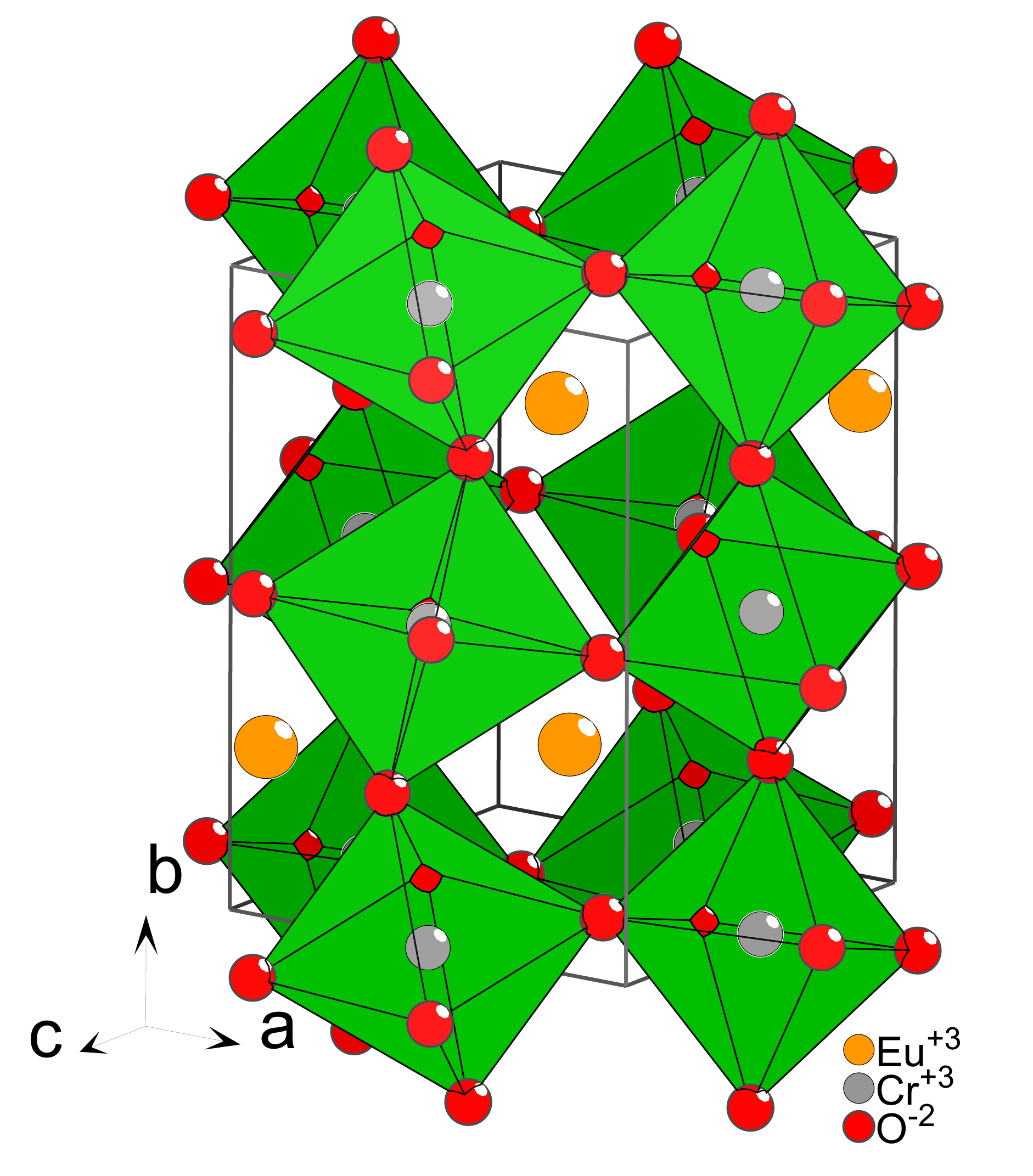}
	\caption{(color online) Crystal structure of EuCrO$_3$. Orange, red, and gray circles represent europium, oxygen and chromium atoms, respectively. Oxgen atoms connect neighboring CrO$_6$ octahedra via common corners.
Structural parameters were taken from the x-ray data which are compiled in Table \ref{Table2}. The lattice parameters at room temperature amount to 5.346 {\AA}, 5.511 {\AA}, and 7.629 {\AA} for $a$, $b$, and $c$, respectively.
}
	\label{Fig1}
\end{figure}

The RCrO$_3$ compounds crystallize with the orthorhombic GdFeO$_3$ structure-type (centrosymmetric spacegroup $Pbnm$) which is highlighted in Fig. \ref{Fig1}. Bertaut and collaborators analyzed the possible symmetry-compatible ordered magnetic structures of the orthorhombic ABO$_3$ compounds.\cite{Bertaut1967,Smart3B} With respect to the possible symmetry adapted modes for the Cr atoms, four  representations of the base vectors are possible, three of them allow a weak ferromagnetic component in the presence of anisotropic exchange forces.

Magnetoelectric effects as well as magnetic and electric field induced switching of the dielectric polarization have been detected in SmCrO$_3$, GdCrO$_3$, and ErCrO$_3$.\cite{Rajeswaran2012}
Polarization in ErCrO$_3$ disappears below the spin orientation transition of $\sim$ 22 K which was ascribed to the crucial connection of the magnetoelectric behavior to the weak ferromagnetism.\cite{Rajeswaran2012, Holmes1970} This conclusion was subsequently questioned by Preethi Meher \textit{et al.} who observed qualitatively similar results but weaker in magnitude effects in LuCrO$_3$ with a diamagnetic R constituent.\cite{Meher2014}
At present, there are also diverging opinions about whether the ferroelectric state as evidence by polarization as well as pyroelectric currents occur at the N\'eel temperature as concluded by Rajeswaran \textit{et al.} or whether a ferroelectric state is formed already at higher temperatures as proposed earlier by Lal \textit{et al.} or Sahu \textit{et al.}.\cite{Lal1989,Sahu2007}
For example, Prado Gonjal \textit{et al.} could not identify a strong correlation between the magnetic and dielectric properties of RCrO$_3$ which were prepared by a microwave assisted sintering.\cite{Gonjal2013}

The properties of rare earth orthochromites with most of the rare earth elements have been extensively investigated.  It appears  that the knowledge of the magnetic properties of the system EuCrO$_3$ is limited to magnetization studies and a first Raman scattering investigation carried out down to 90 K.\cite{Tsushima1969,Venkata2013} Preliminary measurements of the pyroelectric and dielectric properties of EuCrO$_3$ have been carried out for $T >$ 300 K.\cite{Lal1989}
The magnetization studies found antiferromagnetic ordering below $\sim$175 K and weak ferromagnetism with a small ferromagnetic moment of $\sim$0.1 $\mu_{\rm B}$ along the $c$-axis.\cite{Tsushima1969,Gonjal2013}. The Raman data evidenced anomalies in the shift of some Raman modes near the N\'eel temperature.\cite{Venkata2013}
Neither  the magnetic structure nor a detailed investigation of the magnetoelastic coupling  or the thermal properties of EuCrO$_3$ have been carried out, so far.
In a preceding publication, we described the preparation, characterization and magnetization of EuCrO$_3$ and CeCrO$_3$  prepared by a solution combustion method and found magnetization irreversibility and exchange bias effects on such samples.\cite{Maryam} In the present paper, we report on the results of a comprehensive investigation of the magnetic and thermal properties of EuCrO$_3$ from room temperature down to liquid helium temperatures. Especially, we performed detailed neutron diffraction experiments using neutrons of different wavelengths and solved the magnetic structure of EuCrO$_3$ conclusively.
We confirm weak ferromagnetism due to weak canting of the Cr moments. Magnetoelastic effects are investigated by low-temperature thermal expansion and x-ray powder diffraction and detailed Raman scattering experiments. Finally, we report on a first characterization of the dielectric properties of EuCrO$_3$ by frequency dependent permittivity measurements.

\section{Experiment}
High-purity powder samples of EuCrO$_3$ were synthesized by the solution combustion method starting from equimolar solutions of high purity europium nitrate, chromium nitrate and glycine.
 The ground powder was first calcinated at  500 $^\circ$C for 5 h and reground and sintered at  950 $^\circ$C for 12 h.
process.The products were characterized with respect to the composition,  particle size and phase purity by x-ray diffraction as described in detail elsewhere.\cite{Maryam}

All neutron powder diffraction (NPD) patterns were collected on 12 g sample powder packed in an Al container using the DUALSPEC C2 powder neutron diffractometer, equipped with a curved 800-wire BF$_3$ detector at the NRC Canadian Neutron Beam Center CNBC, Canada. Two different wavelengths of 2.37 {\AA} and 1.3282 {\AA} wavelengths from the silicon 311 and the 531 reflection, respectively, were used.  A graphite filter was placed in the incident beam in front of the sample to eliminate higher order contributions.
Diffraction patterns collected with the short-wavelength neutrons were employed to obtain Bragg reflections at high $Q$ values to enable reliable refinements of nuclear structure whereas the longer wavelength was used to explore the small $Q$ regime and to search for magnetic scattering.
Additional neutron scattering was carried out at the N5 triple axis spectrometer at the NRC Canadian Neutron Beam Center to trace the temperature dependence of the order paramater in detail. The initial and final neutron energies both were chosen by the pyrolytic graphite PG002 reflections.
In order to reduce the very high absorption cross section for thermal neutrons from the $^{151}$Eu isotope in the natural isotope composition of Eu,  we utilized a flat geometry thin-walled aluminum sample holder which  was long enough to cover the full beam height but  reduced the thickness of the sample to $\sim$ 1 mm.\cite{Ryan}
To trace magnetoelastic coupling effects low-temperature x-ray powder diffraction (XRPD) patterns were collected with a Bruker D8 Discovery System equipped with an Oxford Instruments closed cycle cooler flat-plate stage.
All profile refinements of the NPD and XRPD patterns were performed with the $FullProf$ software.\cite{FullProf}
DC magnetization measurements were performed using a Magnetic Properties Measurement System (Quantum Design, MPMS). Heat capacity measurements were done in a Physical Properties Measurement System (Quantum Design, PPMS).
Dielectric properties of rectangular shaped samples (6.4 $\times$ 1.6 mm$^2$;  thickness  $\sim$ 0.88 mm) were measured versus temperature at discrete frequencies with an  AH2700A capacitance bridge (Andeen-Hagerling, Inc.). The complex permittivity was analyzed assuming a parallel circuit of a lossless capacitor and a resistor. The linear thermal expansion of a sample of 0.801(1) mm length was measured using a high-resolution miniature dilatometer.\cite{Rotter1998,Reuvekamp2014a,Reuvekamp2014b,Reuvekamp2015}
The Raman spectra were collected on a Jobin Yvon Typ V 010 LabRAM
single grating spectrometer with $\sim$1 cm$^{-1}$ spectral resolution. The spectrometer setup
included a double super razor edge filter, Peltier cooled CCD
camera and a Mikrocryo cryostat with a copper cold finger. Measurements were
performed with linearly polarized He/Ne gas laser light of 632.817 nm with $<$
1mW of power. The light beam was focused to a 10 $\mu$m spot on the top surface of
the sample using a microscope. Measurements were taken in temperatures ranging
between 10 K and 325 K.

\section{Results and Discussion}

\subsection{Chemical and Structural Characterization}
Chemical and structural properties and the morphology  of the samples used for the subsequent investigation have been described in detail elsewhere.\cite{Maryam} Phase purity of the samples was confirmed using room temperature XRPD. Transmission electron microscopy images revealed a uniform size distribution of the particles ranging between 50 - 70 nm. A random particle size distribution ranging between 60 and 80 nm also allowed to model the observed broadening of the Bragg reflections in the XRPD patterns.

\begin{figure}[htp]
	\centering
	\includegraphics[width=8.5cm]{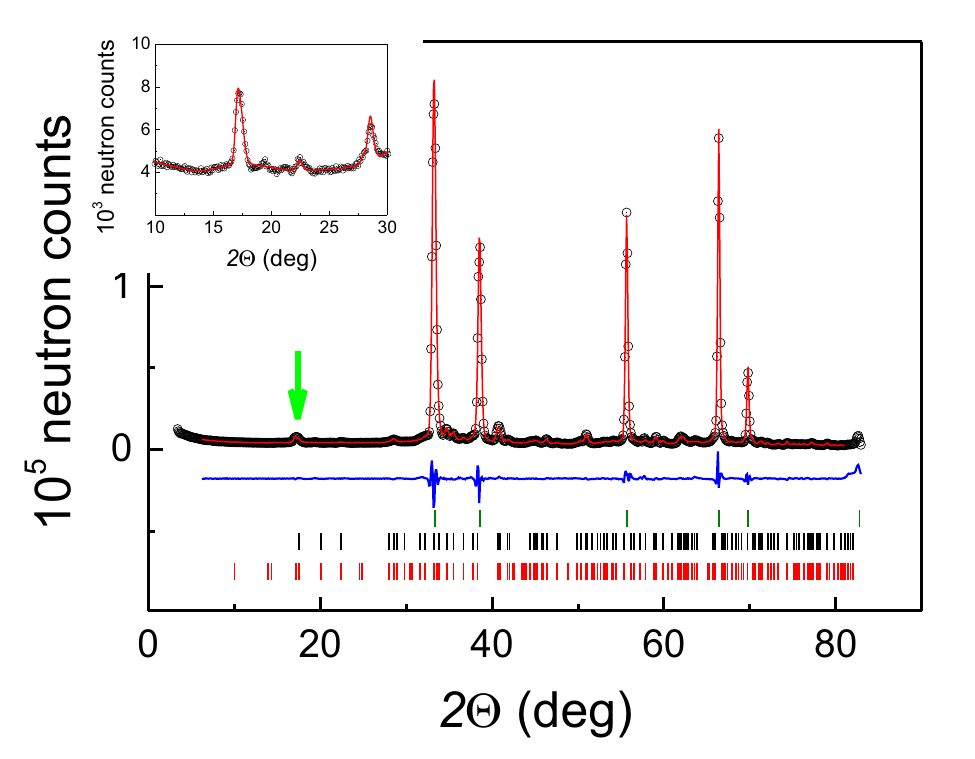}
	\caption{(color online) The NPD pattern of EuCrO$_3$ collected at 3.5 K with $\lambda$ = 1.3282{\AA}. The black circles represent the measured data, the red solid line is the result of the profile refinement. The blue solid line at the bottom of the graph shows the difference between the observed and calculated patterns. The vertical tics (olive, black, red) mark the angles of the Bragg reflections used to simulate the refined pattern for Al, nuclear and magnetic scattering, respectively. The vertical green arrow shows the (011)/(101) magnetic reflections near 17.2$^{\rm o}$ ($d$ = 4.41(2) {\AA}) highlighted in the upper left inset.}
	\label{MagNeutPatt}
\end{figure}

Fig. \ref{MagNeutPatt} displays an NPD pattern of EuCrO$_3$ collected with a neutron wavelength of 1.3282 {\AA} at  3.5 K. An analogous pattern was collected at 280 K (not shown). Both  patterns contain dominant reflections from the Al sample container which was used to shape a thin sample in order to reduce neutron absorption of the Eu isotope $^{151}$Eu. In addition to the nuclear scattering of  EuCrO$_3$, the 3.5 K NPD pattern reveals magnetic scattering with a dominant magnetic Bragg reflection at $d$ = 4.41(2) {\AA}. For the profile refinement of the NPD patterns, we first refined the Bragg reflections of the Al container by partially treating them as special reflections allowing individual shifts and widths. Subsequently, the EuCrO$_3$ nuclear pattern was refined by varying the lattice parameters, atom positions, and the thermal displacement parameters. The background was modeled by a Tschebyscheff polynomial of higher degree. The zero-point of the pattern was also refined but it remained below 0.1$^{\rm o}$. The nuclear structural parameters of the refined NPD patterns at 280 and 3.5 K as well as structural parameters obtained from room temperature XRPD patterns, are compiled in Table \ref{Table2}.

The positional parameters of all atoms generally agree  well with those obtained from the XRPD patterns \cite{Maryam}, with a slight discrepancy seen for the Eu atom positions resulting from the 280 K pattern.
Especially, the NPD data confirm the atom positional parameters of the oxygen atoms O1 and O2 derived from the XRPD data. All cell parameters exhibit a decrease by lowering the temperature. The relative decrease is most pronounced in the $c$ lattice parameter. A shift of the oxygen atoms could not be detected upon lowering the temperature to 3.5 K.  Furthermore, indication for a  structural phase transition between 280 and 3.5 K was not found from the NPD patterns.

\begin{table}[htp]
\caption{Structural parameters and conventional reliability indicators of EuCrO$_3$ as obtained from the Rietveld profile refinement (\textit{FullProf}) of the XRPD pattern collected with Mo-$K_{\alpha 1}$ radiation at $T$ = 295 K and NPD patterns using $\lambda$ = 1.3282 {\AA}\ at $T$ = 3.5 K and 280 K, respectively. The profile refinements were performed within the space group $Pbnm$ (no. 62). The respective site occupancies were not refined.  An absorption correction of $\mu R \sim$ 1 cm and 1.5 cm was assumed for the XRPD and the NPD refinements, respectively \cite{Argonne}. If no error bars are given the parameters have been fixed in the refinements.}
\label{Table2}
\begin{ruledtabular}
\begin{tabular}{c c  c c }
 T (K)  &   295 (XRPD) &   280 K (NPD) & 3.5 (NPD)\\
\hline
a  ({\AA}) &5.34622(9) &  5.3364(9) & 5.3197(8) \\
b  ({\AA})  &  5.51116(8) & 5.4988(9) & 5.4927(7)   \\
c  ({\AA})  & 7.62931(13) &  7.6122(13) & 7.5936(11) \\
V  ({\AA})$^3$& 224.79(1)& 223.37(7) & 221.85(6)\\
& &  &\\
Eu  (4c) &  & &   \\
x & -0.01313(17) & -0.0094(39)& -0.0062(39) \\
y &0.05413(9) &  0.0400(26) & 0.0537(26) \\
z  &1/4  &  1/4 & 1/4\\
$B_{iso}$ ({\AA}$^2$)  &0.140(11)  & 0.54(27) & 0.48(28)\\
& & &\\
Cr (4b) &  &  &  \\
x &1/2  &  1/2 & 1/2 \\
y & 0 &  0 & 0  \\
z & 0 &  0 & 0 \\
$B_{iso}$ ({\AA}$^2$)  & 0.18(3) & 0.63(37) & 0.55(38) \\
& & &\\
O1 (4c) &  & &   \\
x &0.094(2)  & 0.1053(37) & 0.0999(36)  \\
y &  0.460(1) &  0.4746(38) & 0.4785(32)  \\
z & 1/4 & 1/4  & 1/4 \\
$B_{iso}$ ({\AA}$^2$)  &0.7  & 1.0(3) & 1.0(3) \\
& & \\
O2 (8d) &  & &   \\
x &-0.2886(13)  & -0.2884(28) & -0.2953(25)   \\
y & 0.2987(12) &  0.2863(26) & 0.2863(23) \\
z &0.0483(10) & 0.0448(16)  & 0.0500(17)  \\
$B_{iso}$ ({\AA}$^2$)  & 0.74(25) &  0.74(25) & 1.0(3)\\
\hline
Bragg R-factor (\%) & 1.72 & 7.9 & 6.4\\
$R_f$-factor (\%) & 1.09 & 4.4 & 4.0\\
\end{tabular}
\end{ruledtabular}
\end{table}

\subsection{Magnetic Properties}
Our magnetic susceptibility data of EuCrO$_3$ show magnetic ordering at about $\sim$ 178(1) K, as
had already been described by Tsushima \textit{et al.}\cite{Tsushima1969}  and recently by Prado-Gonjal \textit{et al.}\cite{Gonjal2013}

\begin{figure}[htp]
	\centering
	\includegraphics[width=8.5cm]{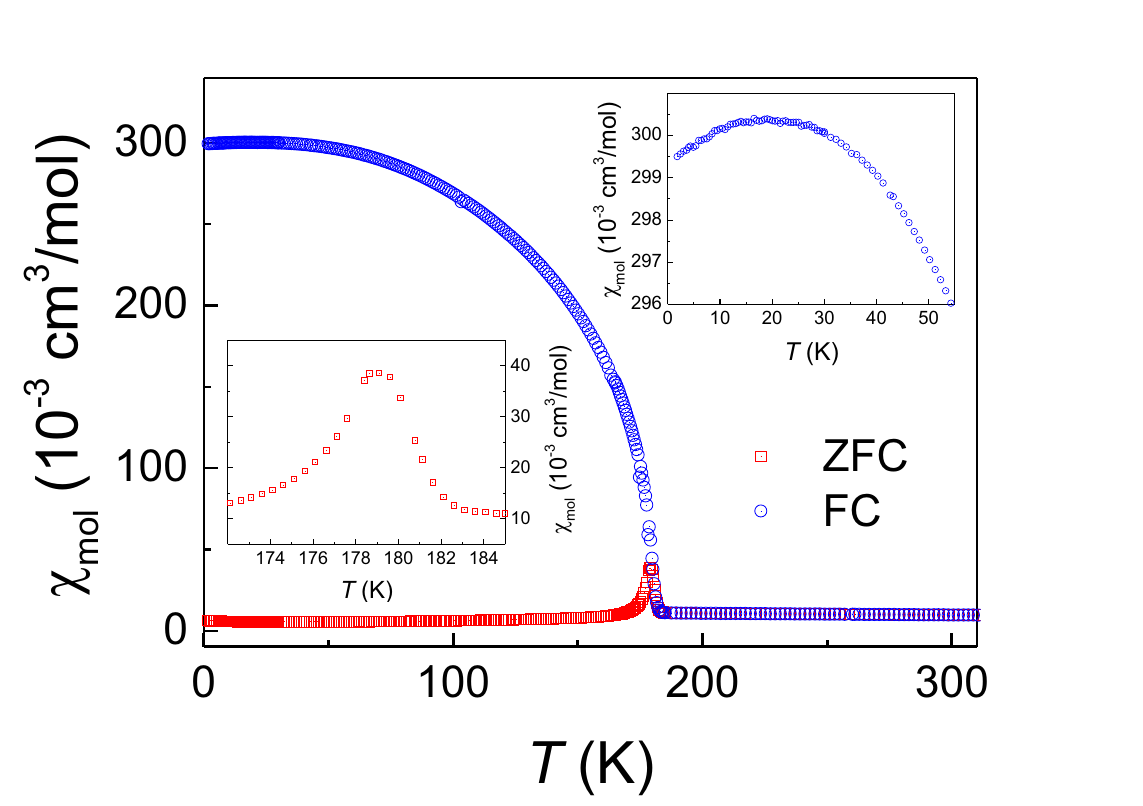}
	\caption{(color online) ZFC (red line) and FC (blue line) magnetic susceptibility determined in a field of 0.05 T of a polycrystalline sample of EuCrO$_3$.}
	\label{Fig3}
\end{figure}

The temperature dependence of the magnetization for EuCrO$_3$  shown in Fig. \ref{Fig3}  exhibits a pronounced thermal hysteresis with a splitting of field cooled (FC) and zero-field cooled (ZFC) susceptibility starting at $\sim$178 K, indicating a weak ferromagnetic component with a saturation of the FC component at low temperatures. This weak ferromagnetic component \cite{Tsushima1969} has been attributed to a slight spin canting of the Cr$^{3+}$ moments.

At sufficiently high temperatures above long-range ordering, the magnetic susceptibility follows a Curie-Weiss law according to
\begin{equation}
\chi_{mol} = \frac{C}{T-\Theta_{\rm CW}}+\chi_{0}.
\label{eqCW}
\end{equation}

The Curie constant, $C$ = $C_{\rm Cr, Eu}$ which depends on the Avogadro number $N_{\rm A}$, the Boltzmann constant $k_{\rm B}$, the $g$-factor $g$, the Bohr magneton $\mu_{\rm B}$ and the spin values of the respective magnetic entities is given by
\begin{equation}
C = N_{\rm A}g^2{\mu_{\rm B}}^2S(S+1)/3k_{\rm B}.
\label{CWconst}
\end{equation}

The term, $\chi_0$ = $\chi_{\rm dia}$ + $\chi_{\rm VV}$  in Eq. (\ref{eqCW}), was added to take care of the temperature independent diamagnetic ($\chi_{\rm dia}$) and   Van Vleck contributions ($\chi_{\rm VV}$).
From the tabulated diamagnetic increments for the individual ions, $\chi_{\rm dia}$ was estimated to contribute -68$\times$10$^{-6}$ cm$^3$/mol to the temperature independent part of the magnetic susceptibility.\cite{Selwood1956}
Van Vleck contributions which become especially important for the magnetism of Eu$^{3+}$ are weakly temperature dependent and will be discussed below.

For  convenience sake, one defines the effective magnetic moment as
\begin{equation}
\mu_{\rm eff}/\mu_{\rm B} = g \sqrt{S(S+1)}.
\label{Mueff}
\end{equation}

For Cr$^{3+}$ with three electrons occupying the $t_{\rm 2g}$ orbitals, in a first approximation, it is justified to assume spin-only magnetism
with $g_{\rm Cr^{3+}} \approx$ 2 and a spin value of $S_{\rm Cr^{3+}}$ = 3/2.\cite{Abragam1970}
Accordingly, one expects an effective magnetic moment for Cr$^{3+}$close to $\mu_{\rm eff}= 3.87 \mu_{\rm B}$.

\begin{figure}[htp]
	\centering
        \includegraphics[width=8.5cm]{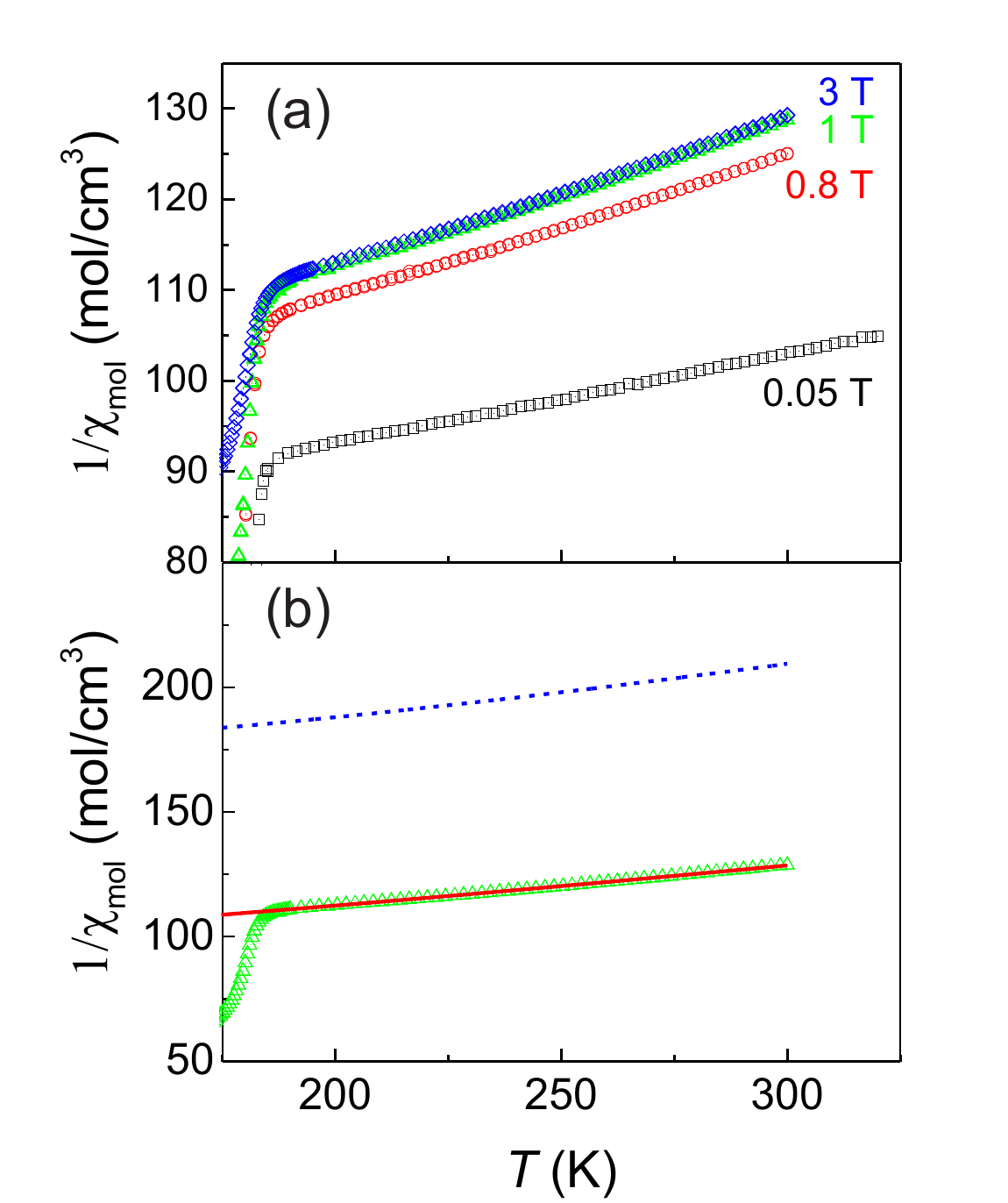}
	\caption{(color online) Inverse magnetic susceptibility of EuCrO$_3$ (a) measured with different applied magnetic fields.  The field dependence arises most likely from a trace ferromagnetic impurity (e.g. EuO or CrO$_2$, $\approx$ 20 ppm) which is gradually saturated with higher fields. (b)  Inverse magnetic susceptibility of EuCrO$_3$ measured with a field of 1 T. The solid red lines represent fits of a Curie-Weiss law (Eq. (\ref{eqCW})) to the data. For the Curie constant we have used the value of 1.873 cm$^3$K/mol corresponding to an effective moment of Cr$^{3+}$ of 3.87 $\mu_{\rm B}$. Blue dashed line: The van Vleck contribution from Eu$^{3+}$ discussed in detail in the text.}
	\label{InvChi}
\end{figure}

Fig. \ref{InvChi} displays the inverse magnetic susceptibilities of EuCrO$_3$ at high temperatures. The data reveal a  magnetic field dependence for small magnetic fields which we attribute to a slight trace of ferromagnetic impurities, possibly EuO or CrO$_2$, which saturates at higher fields. The data collected at 1 T and 3 T fall on top of each other and these are used for further analysis.

A first analysis of the high temperature susceptibility data in terms of the Curie-Weiss law (Eq. (\ref{eqCW})) points to a negative Curie-Weiss temperature about three times larger in magnitude than the long-range ordering temperature. The negative Curie-Weiss temperature indicates sizable predominant antiferromagnetic spin exchange interaction. The derived Curie-Weiss Constant of $C$ $\approx$ 6.1 cm$^3$K/mol gave an effective magnetic moment of 6.98 $\mu_B$. If we attribute the difference to the Cr$^{3+}$ moment simply to the magnetism of the Eu$^{3+}$ ion by using $\mu_{eff}$ = $\sqrt{(\mu_{Cr^{3+}})^2 + (\mu_{Eu^{3+}})^2}$, we arrive at effective moments for Eu$^{3+}$ of the order of 5.9 $\mu_B$, which is by far too large compared to what is expected for Eu$^{3+}$ systems at room temperature ($\mu_{Eu^{3+}}$ = 3.4$\mu_B$ \cite{Earnshaw1968}).

In order to clarify this discrepancy  we followed an alternative approach:
The magnetic moment of Eu$^{3+}$  emerges due to the gradual thermal population of $^7F_J$ states with $J$ taking values of $J$ = 0, 1, ..., 6. Accordingly,  Eu$^{3+}$ typically shows a weak temperature dependent van Vleck  paramagnetism with a typical room temperature value of about 4$\times$10$^{-3}$ cm$^{3}$/mol.\cite{Lueken1999} The  effective magnetic moment derived using Eq. (\ref{Mueff}) exhibits a strong temperature dependence approaching zero at low temperatures.
Using the energy separation of the $^7F_J$ states which is determined by the spin orbit coupling parameter $\zeta_{4f}$ and calculating their thermal population, the effective moment and the susceptibility of Eu$ ^{3+}$ have been derived  e. g. by Lueken.\cite{Lueken1999} $\raisebox{0.25em}{$\chi$}_{VV}$(Eu$^{3+}$) shows a weak temperature dependence and typically increases from liquid He temperatures to room temperature by about 15\%.\cite{Tanner1994}

In order to fit the magnetic susceptibility of EuCrO$_3$ we included the Cr part according to $\raisebox{0.25em}{$\chi$}_{CW}$(Cr$^{3+}$) and allowed in Eq. (\ref{eqCW}) in addition to the temperature independent diamagnetic contributions  a temperature dependent  Van Vleck susceptibility of Eu$^{3+}$, $\raisebox{0.25em}{$\chi$}_{VV}(T)$.
Fig. \ref{InvChi}(b) shows the fit of Eq. (\ref{eqCW}) to the experimental data including diamagnetic, Van Vleck contribution from Eu$^{3+}$ ions and a Curie-Weiss contribution from the Cr$^{3+}$ ions applying a Curie constant $C$ = 1.873  cm$^3$K/mol corresponding to an effective moment of $\mu_{\rm eff}= 3.87 \mu_{\rm B}$.
In order to fit the data, it was found necessary to multiply the Eu contribution by a factor of $\sim$1.26 to enhance the Van Vleck susceptibility. This enhancement can either be due to the magnetic polarization of the Eu moments by the Cr moments or it may be ascribed to a slight change of the susceptibility due to crystal field splitting effects of the excited $^7F_J$ levels which has not taken into consideration in the calculation of the susceptibility of the free Eu$^{3+}$ moments.\cite{Lueken1999} As a consequence of the proposed fitting procedure employed to fit the high-temperature magnetic susceptibility of EuCrO$_3$, the Curie-Weiss temperature is reduced by about $\sim$35\% but it is found still a factor of two larger than the N\'{e}el temperature.

\begin{figure}[htp]
	\centering
	\includegraphics[width=8.5cm]{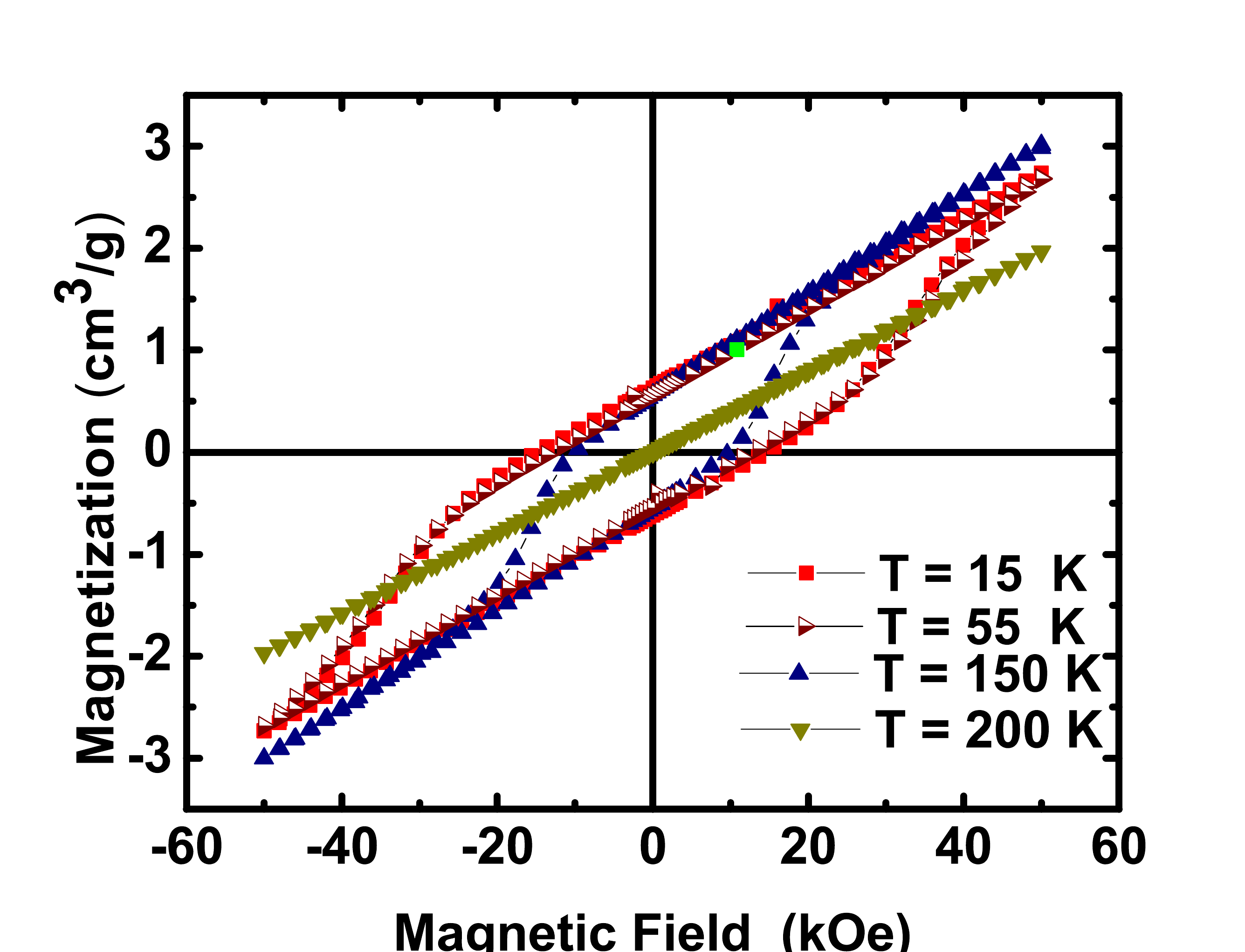}
	\caption{(color online) Gram-magnetization of a polycrystalline sample of EuCrO$_3$ at various temperatures as indicated in the inset.}
	\label{Magnetization}
\end{figure}

Fig. \ref{Magnetization} shows the magnetization versus magnetic field for temperatures above and below the N\'{e}el temperature of 178 K of a polycrystalline sample of EuCrO$_3$. Below the N\'{e}el temperature, a remanent moment is striking which amounts to 0.75 cm$^3$/g or about 0.033 $\mu_{\rm B}$ per Cr atom. If we take  a random orientation of the crystallites into consideration, this finding implies a weak ferromagnetic moment per Cr atom of $\sim$0.1 $\mu_{\rm B}$, in agreement with previous findings.\cite{Tsushima1969}

\subsection{Heat Capacity}
\begin{figure}[htp]
	\centering
	\includegraphics[width=8.5cm]{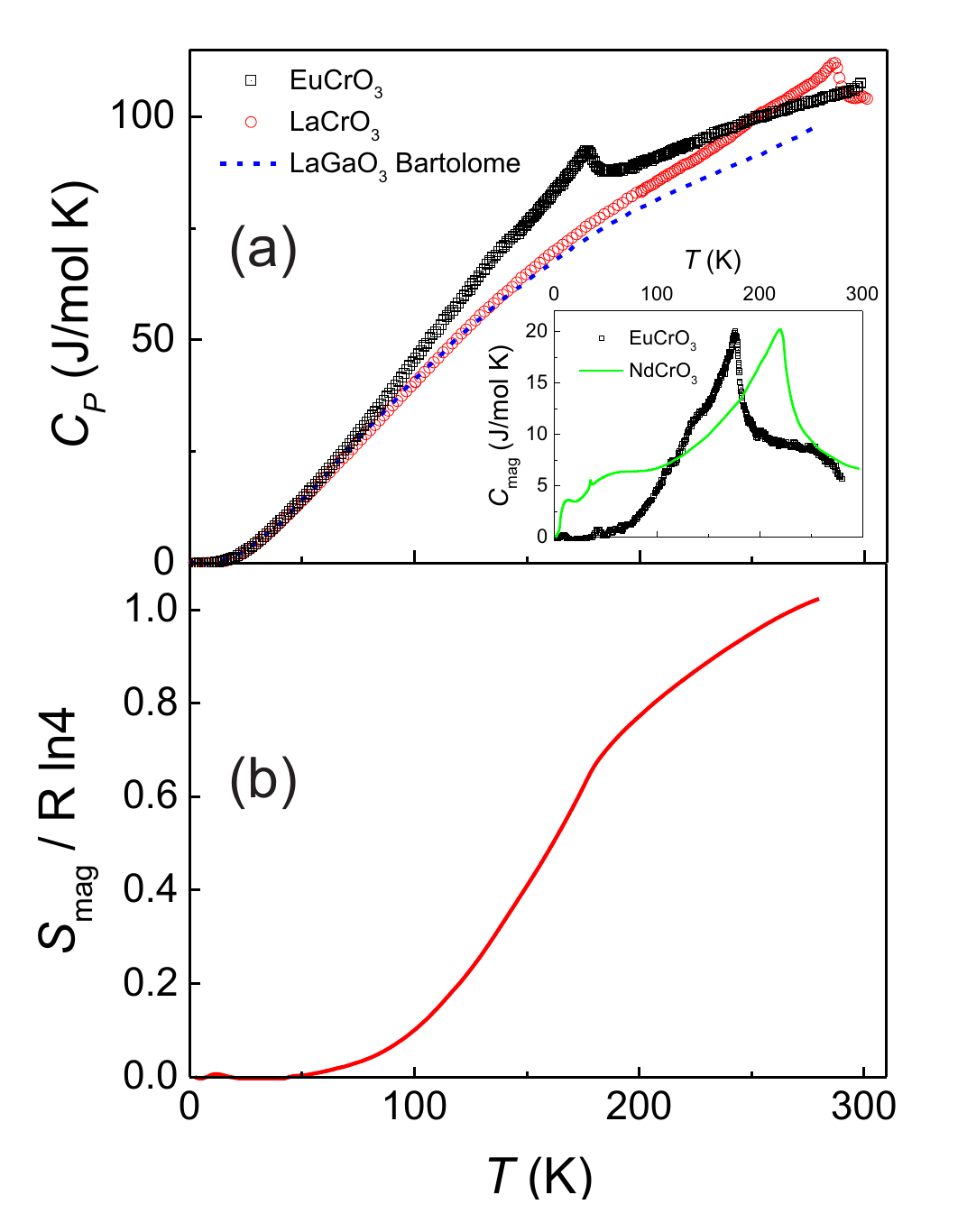}
	\caption{(color online) (a) Specific heat of a nano-powder sample of EuCrO$_3$ (black squares), of LaCrO$_3$ (red circles), and of LaGaO$_3$ (blue dashed line) taken from Bartolom\'{e} \textit{et al.}\cite{Bartolome2000}. The inset shows the magnetic heat capacity of EuCrO$_3$ (black) obtained after subtraction of the lattice contribution (details see text) and of NdCrO$_3$ taken from Ref. \onlinecite{Bartolome2000}, (b) Magnetic entropy of EuCrO$_3$, $S(T)/R$ as obtained according to Eq. (\ref{EqS}). $R$ is the molar gas constant.}
	\label{HeatCap}
\end{figure}

The heat capacities of most rare earth orthochromites, RCrO$_3$ (R = La. Pr, Nd, Sm. Gd, Dy, Ho, Er and Y) have been measured by Satoh \textit{et al.} down to liquid nitrogen temperature and the anomalies associated with the long-range magnetic ordering of Cr$^{3+}$ have been well documented.\cite{Satoh1997} Recently, Bartolom\'{e} \textit{et al.}  and Su \textit{et al.} have extended the heat capacity measurements on the systems NdCrO$_3$, YbCrO$_3$, and HoCrO$_3$ down to liquid helium temperatures.\cite{Bartolome2000,Su2010,Su2011} It appears that only Bartolom\'{e} \textit{et al.} have ventured to compare the heat capacity with a lattice reference (LaGaO$_3$) in order to extract the magnetic contributions to the heat capacity of NdCrO$_3$. They were especially  interested in the crystal field splitting of the Nd$^{3+}$ ions but also worked out the magnetic entropy  associated to the long-range ordering of the Cr moments. To the best of our knowledge, the heat capacity of EuCrO$_3$ has not been determined yet. Fig. \ref{HeatCap}(a) shows the heat capacity of EuCrO$_3$ and that of LaCrO$_3$ and the non-magnetic reference material LaGaO$_3$ for comparison. The latter was taken from the work of Bartomolom\'{e} \textit{et al.}. \cite{Bartolome2000} The magnetic contribution to the heat capacity of EuCrO$_3$, $C_{\rm mag}$, (see inset Fig. \ref{HeatCap}(a)) was obtained by subtracting a lattice reference which was calculated by down-scaling the temperatures of the heat capacity of LaGaO$_3$ by a Lindemann factor.\cite{Lindemann1910,Tari2003} The Lindemann factor which takes into account  the atom mass differences of the constituents of EuCrO$_3$ and LaGaO$_3$ was estimated from the atomic weights and the volumes of the nuclear unit cells to $\sim$ 0.95.\cite{Kim2007} The magnetic contribution to the heat capacity is characterized by a $\lambda$-shaped peak at 176(1) K. The anomaly is somewhat narrower than the analogous anomaly found for NdCrO$_3$.\cite{Bartolome2000} The latter also does not exhibit the shoulder seen in the $C_{\rm mag}$ of EuCrO$_3$ at $\sim$ 130 K.

The magnetic entropy, $S_{\rm mag}$, removed by the long-range ordering of the Cr moments in EuCrO$_3$ was calculated by integrating $C_{\rm mag}/T$, according to
\begin{equation}
S_{\rm mag}(T)=\int_0^T C_{\rm mag}(T')/T' dT'.
\label{EqS}
\end{equation}
At 275 K, the magnetic entropy   amounts to
\begin{equation*}
S_{\rm mag} = R\,{\rm ln}\,(4),
\end{equation*}
in agreement with the entropy
\begin{equation}
S_{\rm mag}=R{\rm ln}(2S+1)=R\,{\rm ln}\,(4),
\label{Smag}
\end{equation}
expected for a $S$ = 3/2 system.

The detailed temperature dependence of the magnetic entropy displayed in Fig. \ref{HeatCap}(b) reveals that about 70\% of the total magnetic entropy are removed below the N\'{e}el temperature. Only 30\% are acquired in the regime of critical fluctuations above $T_{\rm N}$.

Our heat capacity results confirm long-range magnetic ordering below a N\'{e}el temperature of 176 K, in  agreement with the magnetization data. The entropy data pinpoint a $S$ = 3/2 magnetic system ordering consistent with the $d^3$ configuration of the Cr$^3$ cations while contributions to the magnetic heat capacity arising from the Eu$^{3+}$ cations are not apparent.
Bartolom\'{e} \textit{et al.} found a similar result for Cr ordering in NdCrO$_3$, but also detected an additional contribution from Nd crystal electric field excitations below $\sim$100 K.\cite{Bartolome2000}

\subsection{Magnetic Structure}

First evidence for magnetic scattering was gained by comparing the diffraction patterns collected at $T$ = 280 and 3.5 K on the Dualspec diffractometer with neutrons of a wavelength of 1.3282 {\AA} (cf. Fig. \ref{MagNeutPatt}). Apart from the dominating Bragg reflections arising from the Al container, weaker nuclear Bragg reflections are seen. In addition to the nuclear Bragg peaks, a magnetic Bragg reflection was identified at 17.15$^{\rm o}$($d$ = 4.41(2) {\AA}). The 280 K pattern is featureless in this part of the diffraction pattern. Measurements at the larger wavelength (see below) reveal that this reflection consists of two reflections which can be indexed as (011)/(101), the latter one shifted  to higher angles, allowing for an understanding of the asymmetric peak shape noticeable in the 1.3282 {\AA} pattern. The refinement of the magnetic structure will be described in more detail below.

\begin{figure}[htp]
	\centering
	\includegraphics[width=7.5cm]{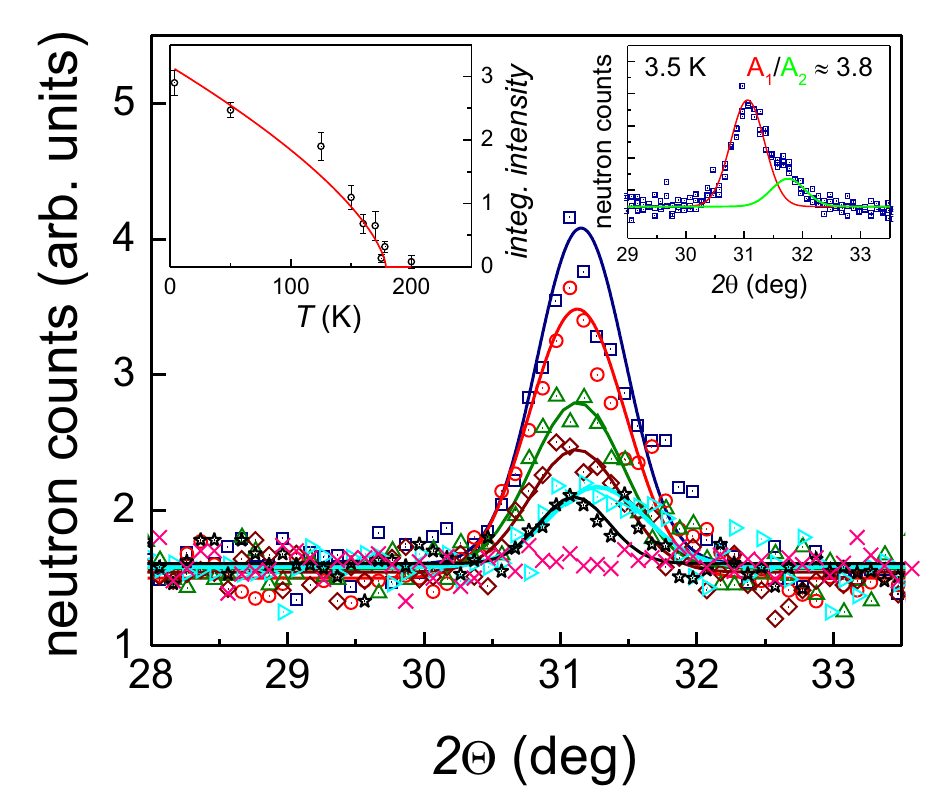}
	\caption{(color online) The (110)/101) magnetic reflection versus temperature collected at a neutron wavelength of 2.37 {\AA}. The solid line is a guide to the eye. The curves refer to data collected at T= 3.5, 50, 125, 150, 160, 170, and 200 K, from top to bottom, respectively. The top right inset displays the decomposition of the diffracted intensity at 3.5 K into two Gaussian peaks with an intensity ratio of 3.8:1. The top left inset shows the temperature dependence of the Gaussian area, (red) A$_1$, of the (011) reflection versus temperature. The solid red line is a fit to a power law with (fixed) critical exponent $\beta \sim$ 0.33 and a critical temperature $T_{\rm c}$ = 178(5) K. }
	\label{MagRef2p37Ang}
\end{figure}

Firstly, we will discuss  the  dependence of the  intensity and the splitting of these reflections on the  temperature. Fig. \ref{MagRef2p37Ang} displays in an overview the temperature dependence of the magnetic reflection observed at $d$ = 4.41(2) {\AA}, highlighting its disappearance between 170 and 200 K. A closer inspection of the shape of this reflection reveals a shoulder on the high angle side, indicating that it is composed of two overlapping reflections. We successfully de-convoluted it by fitting two Gaussian functions, with the Gaussian at the high angle shoulder  having an intensity of about $\frac{1}{4}$ of the main line (see top right inset in Fig. \ref{MagRef2p37Ang}). The splitting of two reflections which amounts to $\sim$ 0.7$^{\rm o}$  is approximately independent of the temperature, indicating temperature stability of the magnetic structure. The two reflections can be indexed as (011) and (101). They are only slightly separated as a consequence of the similar $a$ and $b$ lattice parameters.

\begin{figure}[htp]
	\centering
	\includegraphics[width=7.5cm]{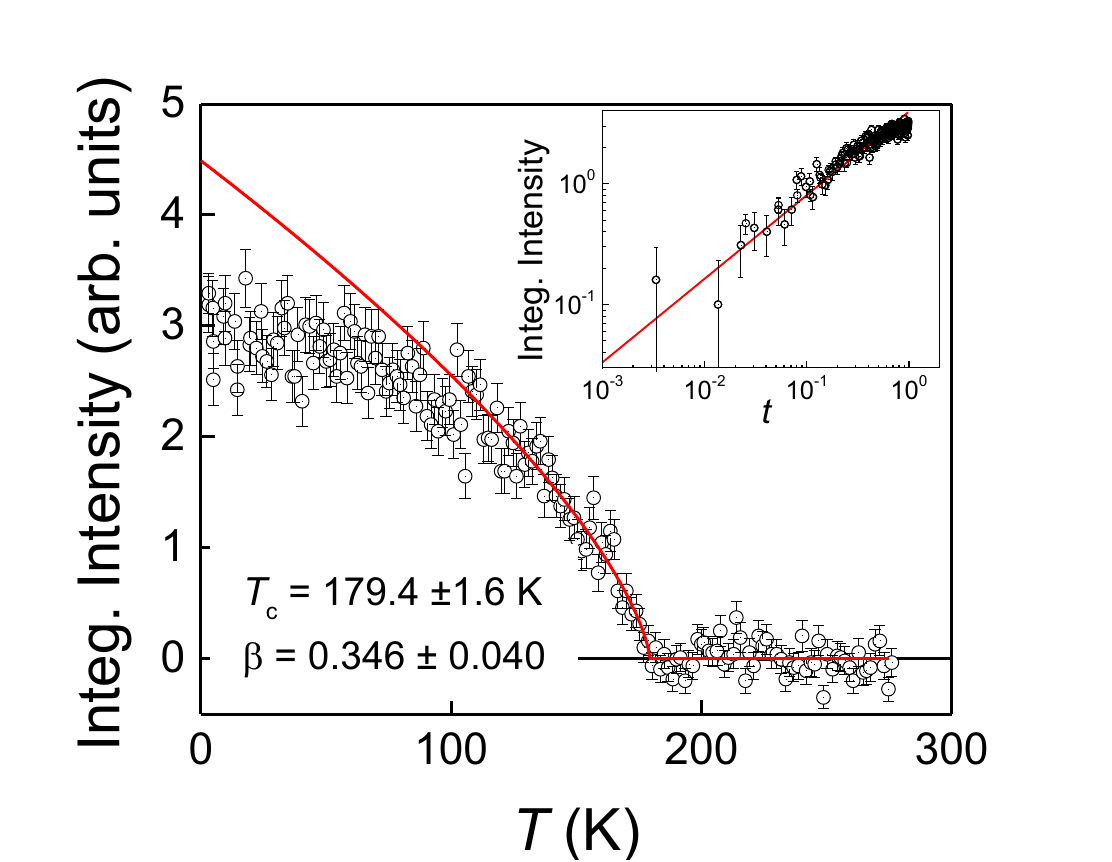}
	\caption{(color online) Integrated intensity of the (011)/101) magnetic reflection of EuCrO$_3$ measured with neutrons of wavelength 1.6513 {\AA} versus temperature. The solid red line represents a fit of a power law according to Eq. (\ref{eqcrit}) with parameters given in the lower inset. The upper inset displays the integrated intensity in a log-log plot versus the reduced temperature, $t$ = ($T$ - $T_{\rm c}$)/$T_{\rm c}$), with $T_{\rm c}$ = 179.4 K. }
	\label{N5PowLaw}
\end{figure}

A more complete temperature dependence of the (011)/(101) magnetic reflection was collected on NRC's  triple-axis spectrometer N5, using an elastic configuration by adjusting $E_{\rm f}$ to 30 meV (1.6513 {\AA}) and the  momentum transfer $Q$ to 1.4164 {\AA}$^{-1}$. Fig. \ref{N5PowLaw} displays the integrated intensity versus temperature in a linear and log-log scale versus reduced temperature.

A power law according to
\begin{equation}
M(T) = M_0 (1-T/T_{\rm c})^{\beta},
\label{eqcrit}
\end{equation}

\noindent where the integrated intensity $\propto$ $M^2$, was fitted to the background-corrected intensities by using data within the reduced temperature range  $t$ = ($T$ - $T_{\rm c}$)/$T_{\rm c}$) $<$ 0.25. The fits resulted  in a critical temperature

\begin{equation}
T_{\rm c} = 179.4 \pm 1.3 {\rm K},
\end{equation}
\noindent which  is good in agreement with the susceptibility and heat capacity data. The critical exponent derived from the fit amounts to

\begin{equation}
\beta = 0.364 \pm 0.010,
\label{critexpo}
\end{equation}

\noindent consistent with the critical exponents of standard universality classes.\cite{LeGuillou1977}
The inset in Fig. \ref{N5PowLaw} shows the data in a log-log scale highlighting the good agreement of the critical power law down to reduced temperatures of 3$\times$10$^{-3}$ with a critical exponent given in Eq. (\ref{critexpo}).

\begin{figure}[htp]
	\centering
	\includegraphics[width=5.0cm]{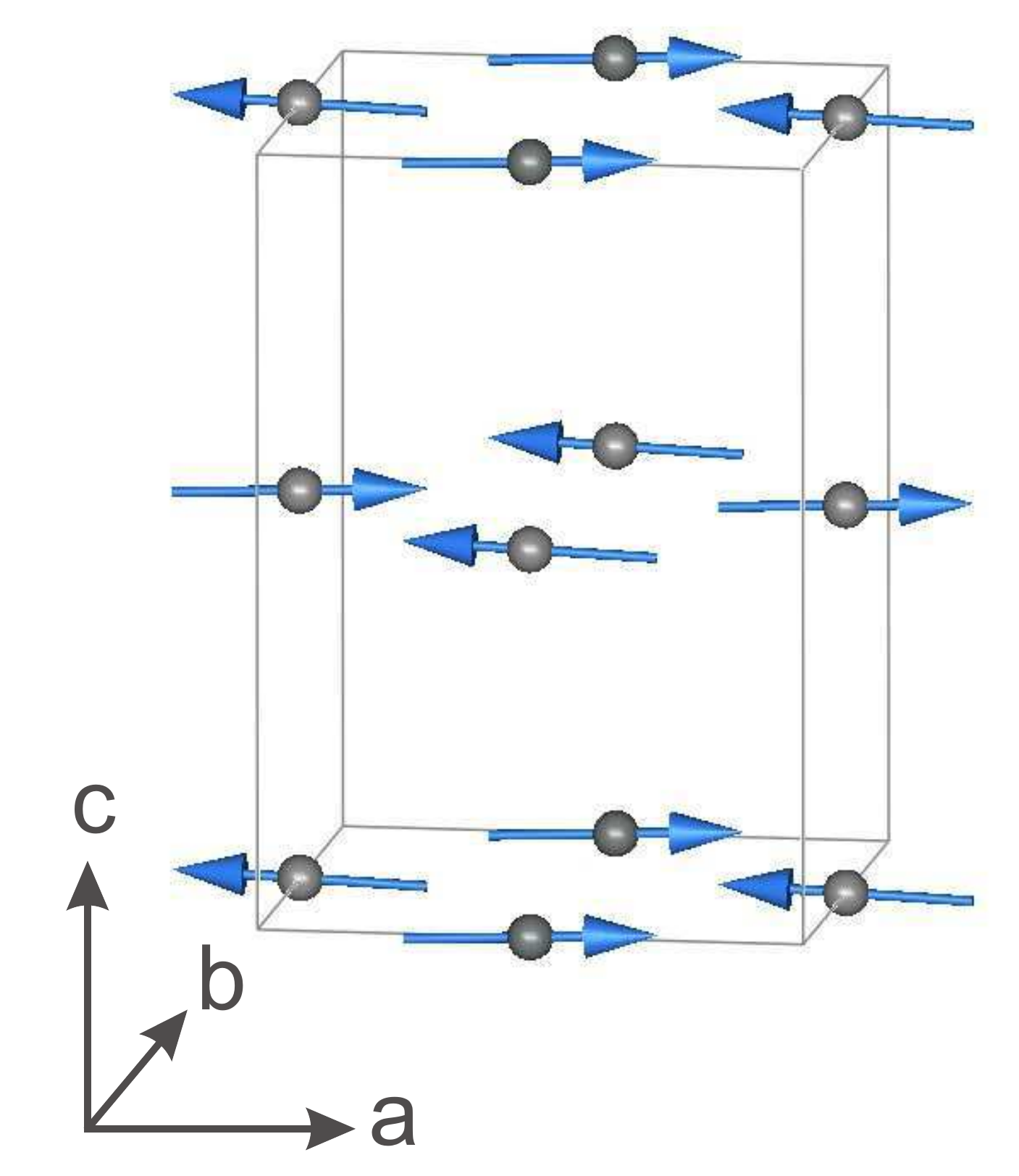}
	\caption{(color online) Magnetic structure of the Cr moments in EuCr$_3$ as refined from the NPD diffraction data. The Cr moments order
 with a $G_x$- type antiferromagnetic structure and a slight canting with a $F_z$ component of 0.1 $\mu_{\rm B}$ implying a tilting angle of the moments with the  $a$ - $b$ plane of $\sim$ 2.4 $^o$.}
	\label{MagStr}
\end{figure}

Finally, we describe the magnetic structure of EuCrO$_3$. The symmetry analysis for orthorhombic ABO$_3$ compounds has been carried out and described in detail by Bertaut.\cite{Smart3B} According to the symmetry analysis, four representations of the base vectors  are possible for the Cr atom, occupying the sites ($\frac{1}{2}$,0,0), ($\frac{1}{2}$,0,$\frac{1}{2}$) (0,$\frac{1}{2}$,$\frac{1}{2}$), and (0,$\frac{1}{2}$,0). Three of these representations, ($\Gamma_3$, $\Gamma_3$, and $\Gamma_4$), are compatible with a weak ferromagnetic component.
A good fit of magnetic refinement of our NPD pattern collected at 3.5 K is achieved based on the $\Gamma_4$ representation, i.e. a $G_x$ magnetic structure (see Fig. \ref{MagStr}) with a negligible component along the $y$ direction. The small but finite ferromagnetic component ($F_z$) which was already indicated by a spontaneous magnetization component in the susceptibility data, is discussed below.

The refined magnetic moment of the Cr atoms at  3.5 K amounts to
\begin{equation}
\mu_x = 2.4 \pm 0.1 \mu_{\rm B},
\label{magmoment}
\end{equation}

\noindent consistent with the expected value of 3$\mu_{\rm B}$ for a $S$= 3/2 system with a $g$-factor $g \sim$ 2.

Fig. \ref{MagStr} displays the magnetic structure of EuCrO$_3$ as refined assuming a dominant $G_x$ spin configuration and a slight canting leading to  a weak ferromagnetic moment along $c$ of 0.1 $\mu_{\rm B}$.

\begin{figure}[htp]
	\centering
	\includegraphics[width=7.5cm]{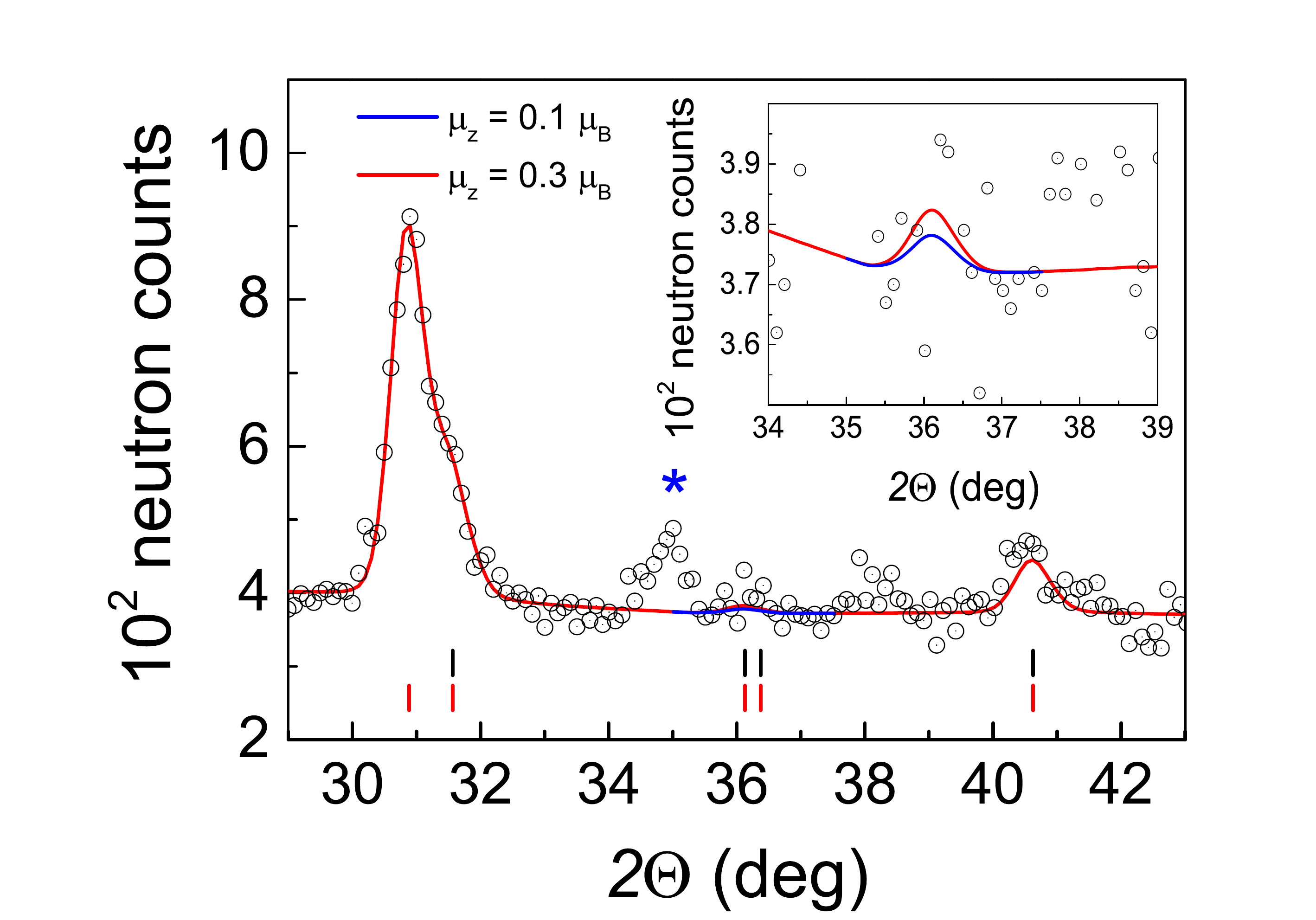}
	\caption{(color online) NPD pattern of EuCrO$_3$ collected at 3.5 K with $\lambda$ = 2.37{\AA}. The (black) circles represent the measured data, the (blue, red) solid lines are the result of the profile refinement (nuclear part and magnetic part) with a $G_x$ antiferromagnetic structure and with a weak ferromagnetic component along $c$ fixed to 0.1 $\mu_{\rm B}$ and 0.3 $\mu_{\rm B}$. The  vertical (black, red) bars mark the angles of the Bragg reflections used to simulate the refined pattern for the nuclear and magnetic scattering, respectively. The (blue) asterisk tags an impurity reflection.}
	\label{SpinCanting}
\end{figure}

In the diffraction pattern collected with $\lambda$ = 1.3282 {\AA}, the (011)/(101) magnetic reflections are overlapping. Choosing a wavelength of 2.37{\AA} allows us to resolve them and to compare their intensities with the expected intensities of the $G_x$ magnetic structure. Fig. \ref{SpinCanting} displays the profile refinement as well as a very weak peak at 36.2$^{\rm o}$. This peak results from the nuclear and magnetic diffraction, the latter is due to the small ferromagnetic component along the $c$-axis. Model calculations (see Fig. \ref{SpinCanting}) indicate the ferromagnetic component to lie between 0.1 and 0.3 $\mu_{\rm B}$, consistent with the magnetization measurements (see above). This finding implies a canting angle of the Cr moment out of the $a$-$b$ plane between 2.4 and 7$^{\rm o}$.

\subsection{Structural Properties at Low Temperatures}

\begin{figure}[htp]
	\centering
  	\includegraphics[width=8.0cm]{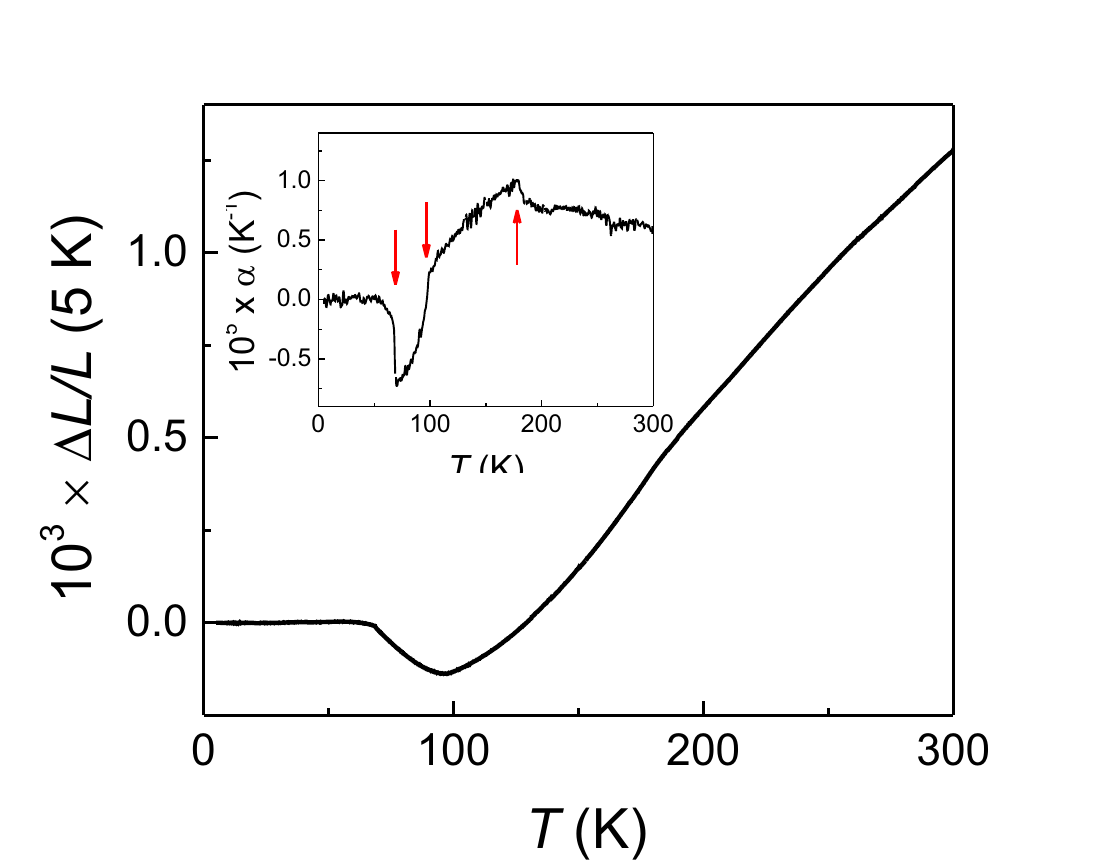}
	\caption{(color online)  Temperature dependence of the length change of a polycrystalline sample of EuCrO$_3$, measured by capacitance dilatometry in a warming cycle. Inset:  Coefficient of the linear thermal expansion, $\alpha(T)$.}
	\label{ThExpa}
\end{figure}

In order to investigate the crystal structure and structural changes below room temperature and to search for magnetoelastic distortions, we have carried out  linear thermal expansion, powder XRPD  and Raman scattering measurements down to liquid helium temperature.  A first overview of thermal and magnetoelastic effects on the lattice properties becomes available from the measurement of the dilatometric length change of a polycrystalline pellet (Fig. \ref{ThExpa}). Starting from lowest temperatures and slowly ($\sim$0.5K/min) raising the temperature, the length change is characterized by an extended plateau whereas the sample starts to contract above  $\sim$ 70 K, i.e. EuCrO$_3$ shows negative thermal expansion between $\sim$ 70 and $\sim$ 100 K. Above $\sim$100K the sample length  grows monotonically   with a slight change of the slope at  $\sim$ 175 K, i.e. in the range of the N\'{e}el temperature. The anomalous behavior below $\sim$100 K  is clearly revealed in the coefficient of linear thermal expansion, $\alpha(T)$ = $d/dT$ ($L(T)/L_{\rm 0}$) (see inset Fig. \ref{ThExpa}) which is negative between $\sim$ 70 and 100 K.

\begin{figure}[htp]
	\centering
  	\includegraphics[width=8.0cm]{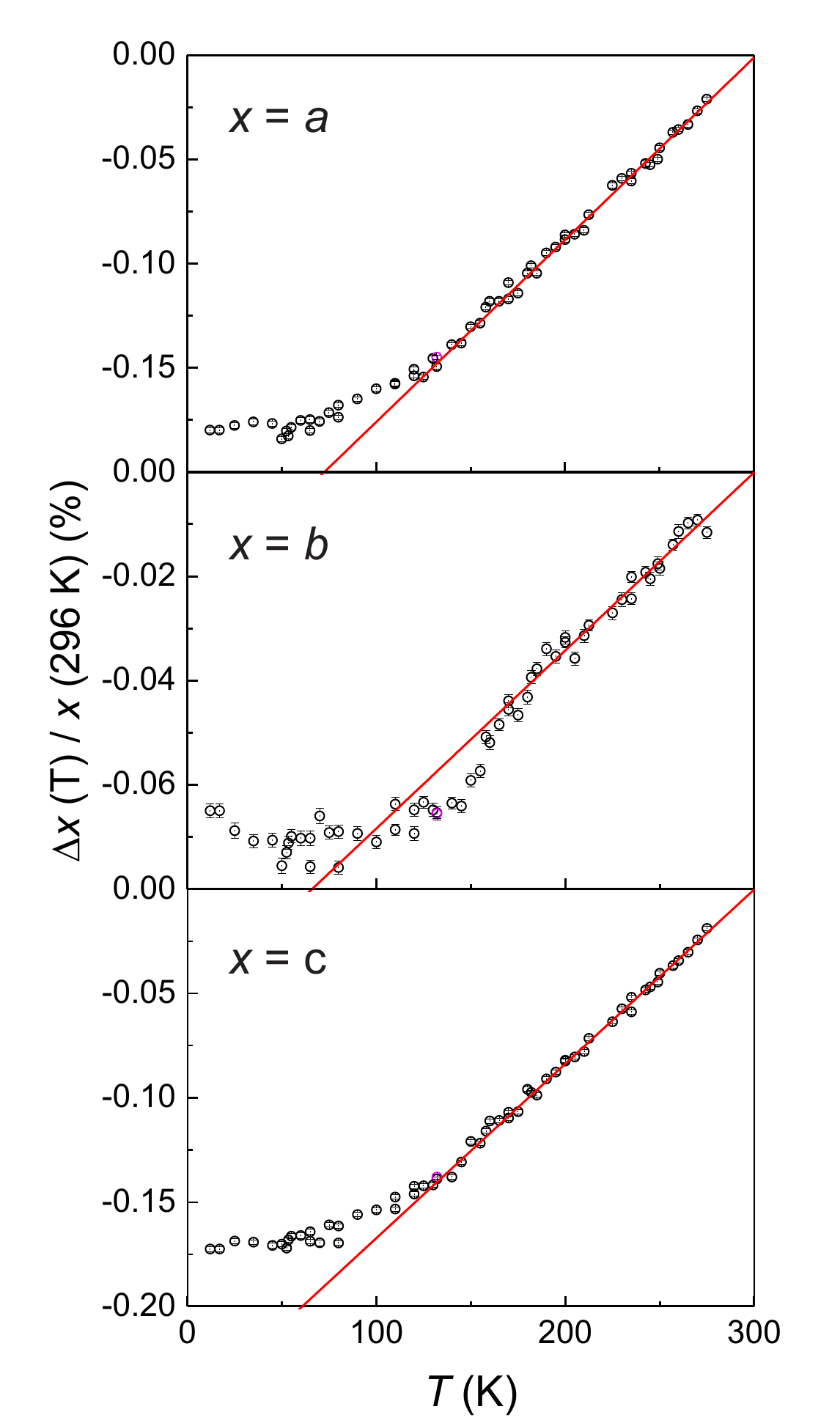}
	\caption{(color online)  Temperature dependence of the lattice parameters of EuCrO$_3$. The (red) solid lines are fits of a linear temperature dependence to the data above 200 K.}
	\label{XRDThExpa}
\end{figure}

The thermal expansion measurements have been carried out on a randomly oriented polycrystalline pellet.
Detailed directional information of the thermal and magnetic effects on the lattice properties was obtained from a series of powder XRPD measurements over the temperature range from 10 K to room temperature. Fig. \ref{XRDThExpa} displays the orthorhombic lattice parameters versus temperature. While $a$ and $c$ show about the same temperature behavior,  the temperature variation of $b$ is by a factor of 2.5 smaller than the contraction of $a$ and $c$. A noticeable anomaly associated with  long-range antiferromagnetic ordering below $\sim$ 178 K, clearly seen in the thermal expansion measurements is only visible in the $b$ lattice parameter. Below $\sim$175 K, $b(T)$ exhibits a slight anomalous decrease below the linear part found for all lattice parameters above  $\sim$ 200 K.  $b$ also exhibits a slight increase towards lowest temperatures, i.e. a slightly negative thermal expansion whereas $a$ and $c$ have a positive thermal expansion throughout the whole temperature range. The cell volume shows  positive thermal expansion in the whole temperature range indicating that the small magnetoelastic distortion in $b$ is compensated by the  temperature behavior of $a$ and $c$.  Subtle anomalies in all XRPD derived lattice parameters between 50 and 70 K may be associated to the negative thermal expansion anomaly seen in the capacitance dilatometer measurements (cf. Fig. \ref{ThExpa}).

Raman scattering investigations on the series of orthochromites RCrO$_3$ ($R$ = Y, La, Nd, Pr, Sm, Gd, Dy, Ho, Er, Yb, Lu)  have been rather extensively carried out at room temperature as well as $\sim$ 100 K in order to assign the phonon modes and to study the lattice response to magnetic ordering either of the Cr or of the rare earth moments.\cite{Iliev2006,Weber2012,ElAmrani2014}
Our dilatometer measurements and the low-temperature XPD data indicated  magnetoelastic distortions of the lattice.
Raman spectroscopy of the phonon modes allows  to trace  very sensitively subtle changes of the lattice and to identify possible structural and magnetic phase transitions.

\begin{figure}[h!]
	\centering
	\includegraphics[width=28.0cm]{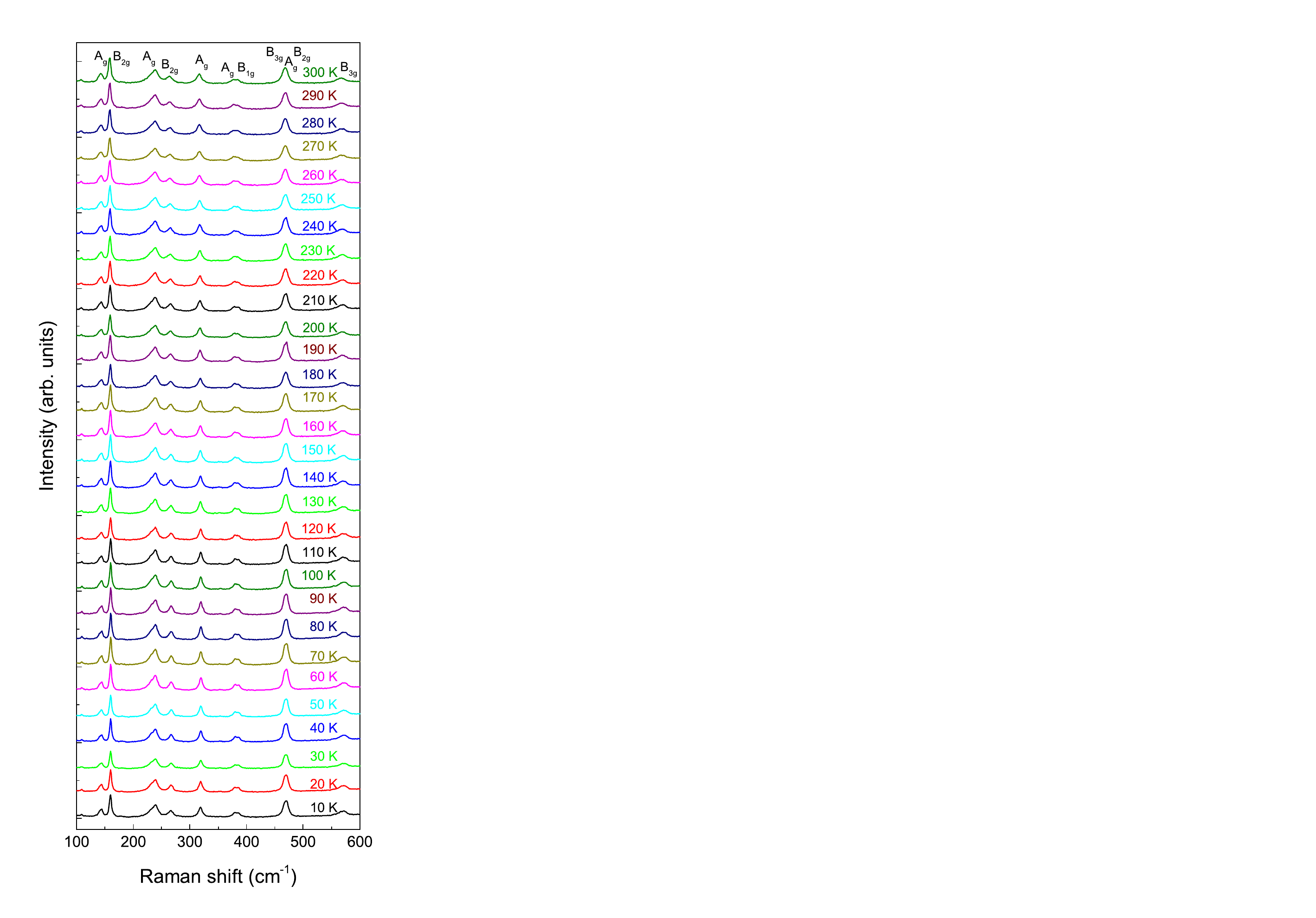}
	\caption{(color online) Raman spectra of EuCrO$_3$ versus temperature. The mode assignment is given above the topmost (300 K) spectrum.}
	\label{Allraman}
\end{figure}

At present,  only one Raman study on EuCrO$_3$ down to $\sim$90 K is available reporting some anomalies near the N\'{e}el temperature in the shift of some modes as well as in their linewidth.\cite{Venkata2013}
Here, we have extended these investigation and examined the Raman spectra down to 10 K and studied in detail the shift of the modes and their linewidth versus temperature.
According to  group theory one expects 24 Raman active modes for EuCrO$_3$: 7 $A_g$ + 5 $B_{1g}$ + 7 $B_{2g}$ + 5 $B_{3g}$, involving vibrations of Eu and the oxygen atoms.\cite{Bilbao} Fig. \ref{Allraman} shows the Raman spectra between 10 K and room temperature with the assignments according to Weber \textit{et al.}.\cite{Weber2012} The spectra contain  only 11 Raman modes, thus the other predicted modes have either too low an intensity to be observed or they are located below our experimental region. The spectra are similar to what has been observed for other rare earth orthochromites and from the dependence of the mode energies on the ionic radii of the rare earth atoms fall between the spectra of GdCrO$_3$ and SmCrO$_3$. The $A_g$ - $B_{1g}$ doublet near 375 cm$^{-1}$ is split whereas the mode at $\sim$ 470 cm$^{-1}$ is still degenerate but clearly broadened as compared to other modes in its neighborhood. Above 600 cm$^{-1}$ a broad band is seen which is centered at 650 cm$^{-1}$. It corresponds to the analogous broad band at $\sim$720 cm$^{-1}$ in LaCrO$_3$ which has been assigned to lattice imperfections possibly involving Cr$^{4+}$ centers.\cite{Iliev2006}
At first glance, all Raman mode frequencies appear to not be affected by lowering the temperature. A closer inspection of the shift and the linewidth of the Raman modes was carried out by fitting Lorentzian lines to the modes.
This procedure discloses especially weak but noticeable softening for all analyzed modes below  $\sim$ 75 K. Some modes show additional magnetic anomalies below  $\sim$ 175 K. Generally, the linewidths decrease and pass through a minimum at  $\sim$ 75 K and grow again towards lowest temperatures. Noticeable magnetic anomalies could not be detected in the mode linewidths at the N\'{e}el temperature around 175 K.

The temperature dependent change of the frequency of a phonon mode of an insulator with frequency $\omega_j(T)$,  comprises contributions from the change of the lattice volume, $\Delta \omega_{j,\rm latt}(T)$ due to thermal expansion or contraction (including lattice anomalies) and  also magnetostrictive induced changes of the unit cell volume, intrinsic multi-phonon anharmonic contributions, $\Delta \omega_{,\rm anh}(T)$, and contributions from spin-phonon coupling effects, $\Delta \omega_{j,\rm sp-ph}$. These  superimpose according to\cite{Granado1999A}

\begin{equation}
\omega(T)_j = \omega_j(0) + \Delta \omega_{j,\rm latt}(T) + \Delta \omega_{j,\rm anh}(T) + \Delta \omega_{j,\rm sp-ph}(T).
\label{TdepModes}
\end{equation}

The lattice contributions  are proportional to the volume change, $\Delta V (T)$ and connected via the mode Gr\"uneisen parameter, $\gamma_j$ via

\begin{equation}
\frac{\Delta \omega_j(T)}{\omega_j (T = 0)} = -\gamma_j \frac{\Delta V(T)}{V(0)}.
\label{Lattice}
\end{equation}

Assuming mode Gr\"{u}neisen parameters $\gamma_j \approx$ 1 our XPD results indicate mode shifts of the order of 0.4\% over the whole temperature range. Especially the magnetoelastic effct seen in the lattice parameter $b$ is very weak and cannot explain magnetic contributions.

Anharmonic decay processes add an additional contribution to the phonon linewidths and lineshifts.
Based on the work of Klemens assuming three and fourfold phonon-decay processes, Balkanski has derived the following relationship for the temperature dependence of the anharmonic line shift and linewidth\cite{Klemens1966,Balkanski1983}

\begin{equation}
\Delta \omega_{\rm anh}(T) = C_2 (1 + 2 n_{\rm BE}(x)) + C_3 (1+ 3 n_{\rm BE}(y) + 3 n_{\rm BE}^2(y)),
\label{Anharm}
\end{equation}

\noindent where $n_{\rm BE}(x)$ and $n_{\rm BE}(y)$ are Bose-Einstein statistical factors, $1/(exp(x)-1)$ and $1/(exp(y)-1)$, respectively, with
$x$ = $\hbar \omega_0/2k_{\rm B}T$ and $y$ = $\hbar \omega_0/3k_{\rm B}T$ appropriate for the decay of an optical phonon into two and three acoustic phonons of identical energy.\cite{Klemens1966} An analogous relationship holds for the phonon linewidth.\cite{Balkanski1983}

Spin-phonon coupling  caused by the modulation of the spin exchange interaction by lattice vibrations adds another channel for
magnetic renormalization of the phonon frequencies.\cite{Baltensperger1968} It is proportional to the spin-spin correlation function $\langle s_i \cdot s_j\rangle$ and the second derivative  of the spin exchange integral $J_{i,j}$ with respect to the coordinate connection the spin $s_i$ and $s_j$.\cite{Granado1999A,Sharma2014}
Using the molecular field approximation of the spin-spin correlation function, $\langle s_i \cdot s_j\rangle \propto M_{\rm sublat}^2(T)$, $\Delta \omega_{\rm sp-ph}(T)$ is proportional to the square of the sublattice magnetization $M_{\rm sublat}^2(T)$ according to

\begin{equation}
\frac{\Delta \omega_{\rm sp-ph}(T)}{\omega_{\rm sp-ph}(0)} \propto \frac{1}{\omega^2(0)}\frac{\partial^2 J}{\partial u^2} M_{\rm sublat}^2(T).
\label{spinphon}
\end{equation}

Spin-phonon coupling renormalization of the mode lineshift is subject to the modulation of the spin exchange and generally effects  modes at different frequencies differently, depending on the manner of how the spin exchange is affected by the relevant lattice vibrations modifying the bonding distances and angles especially to the oxygen cations mediating the superexchange.

\begin{figure}[htp]
\centering
     \includegraphics[width=8.0cm]{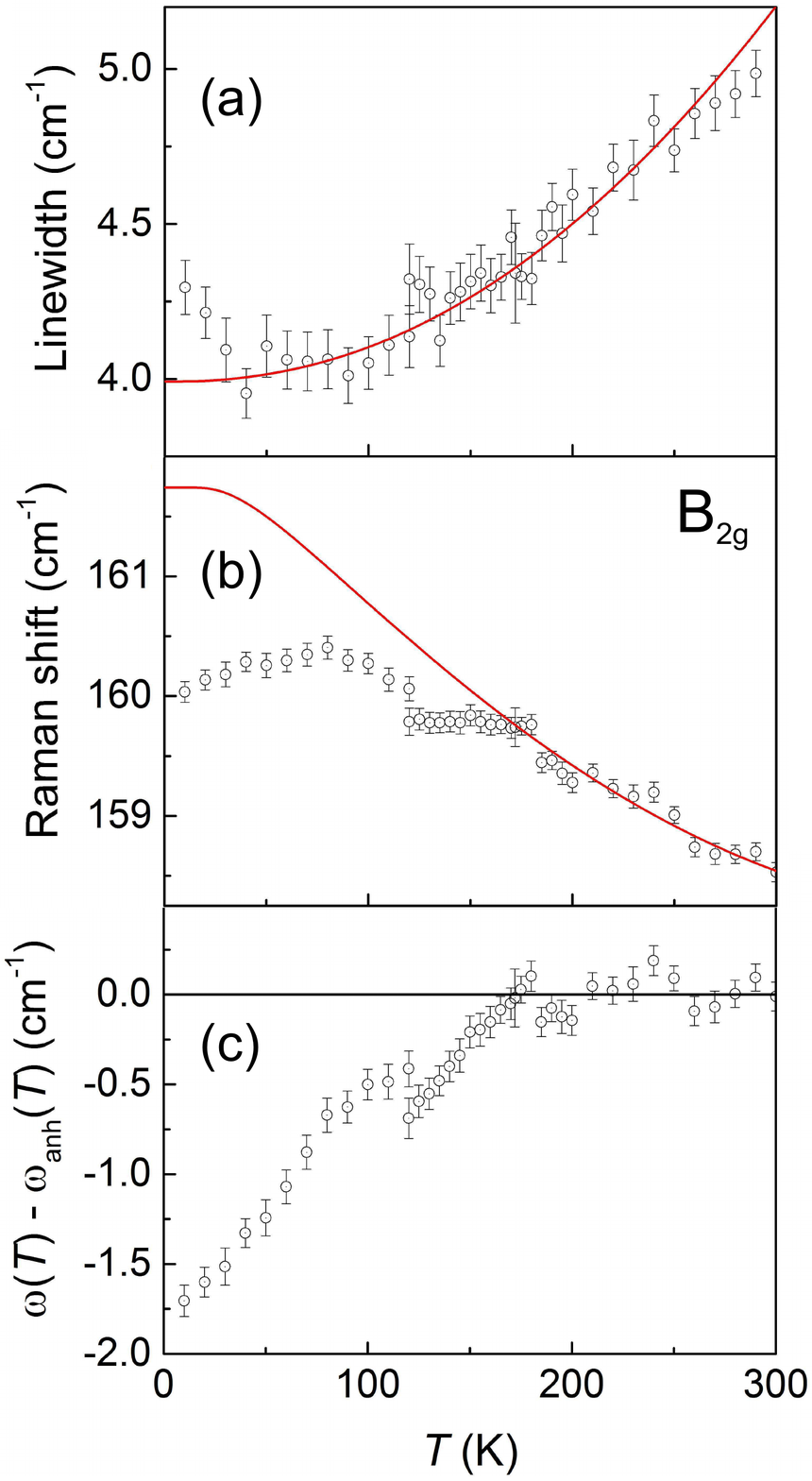}
\caption{(color online) (a) Lorentzian linewidth (FWHM), (b) Raman shift, and (c) the difference $\omega(T)$ - $\omega_{\rm anh}$. The red solid lines were obtained from a fit of Eq. (\ref{Anharm}) to the experimental data taken at temperatures above 200 K and extrapolated to low temperatures. Mode assignment according to Refs.\onlinecite{Iliev2006,Weber2012}.}
\label{Raman160}
\end{figure}

Fig. \ref{Raman160} displays the Raman shift, the linewidth and the residual $\omega(T)$ - $\omega_{\rm anh}$ of the $B_{\rm 2g}$ at $\sim$160 cm$^{-1}$. The mode assignment has been adopted from the Refs. \onlinecite{Iliev2006,Weber2012}.

Raman shift and linewidth above $T \sim$ 175 K follow well a temperature dependence expected for anharmonic phonon decay processes according to Eq. (\ref{Raman160}).
Anomalies in the Raman shift  of the 160 cm$^{-1}$ $B_{2g}$ mode at $\sim$175 K and near $\sim$ correspond well to antiferromagnetic ordering and to the anomaly seen in the thermal expansion (see Fig. \ref{ThExpa}.
Between 175 and 125 K, the Raman shift levels off apparently due to magnetic effects. After a step-like increase below 125 K, the Raman shift passes through a maximum centered at  $\sim$ 75 K and decreases again to lowest temperatures. In contrast to the Raman shift, the linewidth shows no apparent magnetic anomalies in this temperature regime. Though, analogous to the shift, the linewidth attains a minimum at  $\sim$ 75 K and grows again by about $\sim$10\% to the lowest temperatures.

\begin{figure}[htp]
	\centering
	\includegraphics[width=8.0cm]{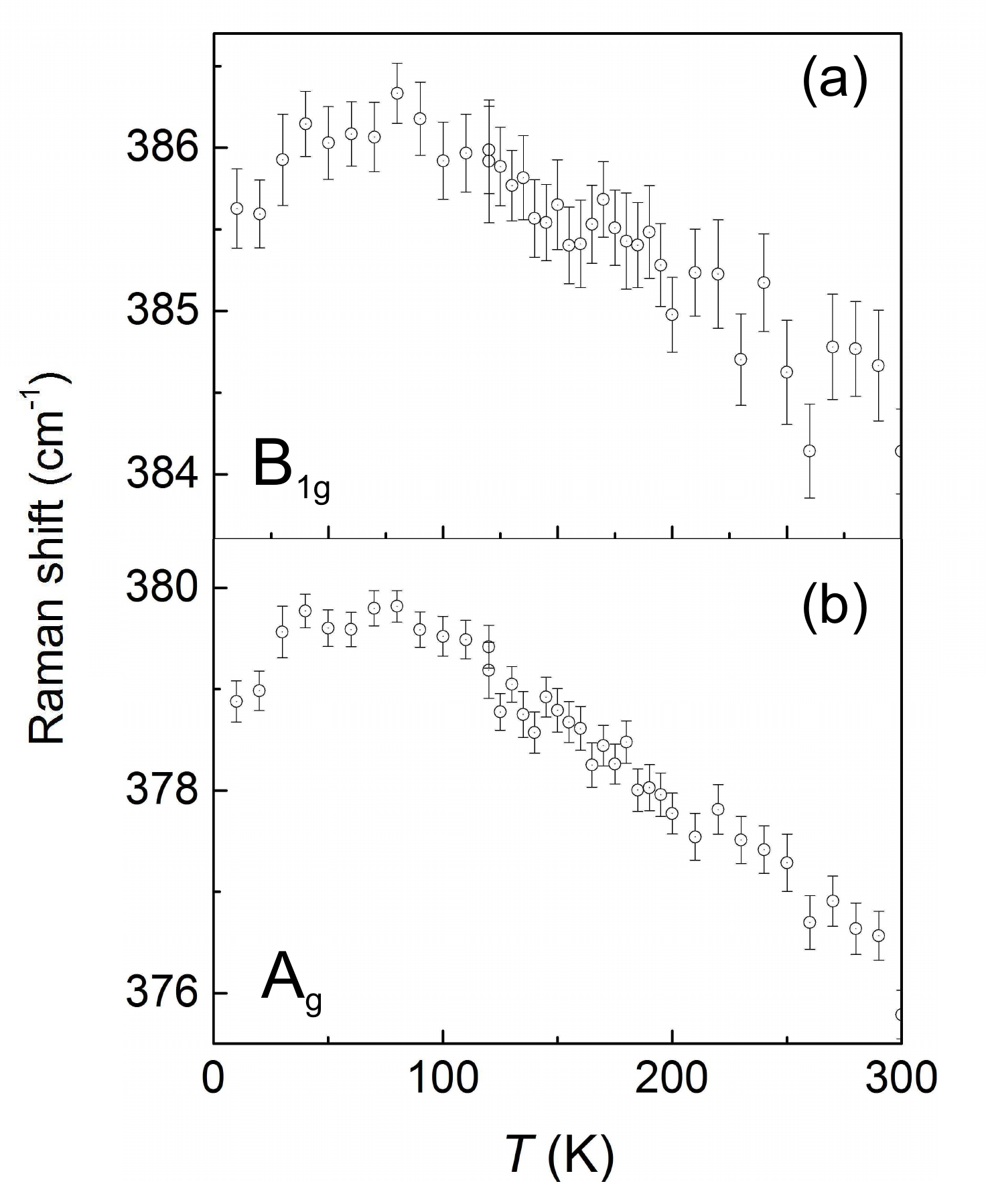}
	\caption{(color online) Raman sifts of the $A_g$ - $B_{1g}$ doublet observed near 380 cm$^{-1}$. Mode assignments according to Refs.\onlinecite{Iliev2006,Weber2012}.}
	\label{Raman1}
\end{figure}

\begin{figure}[htp]
	\centering
	\includegraphics[width=8.0cm]{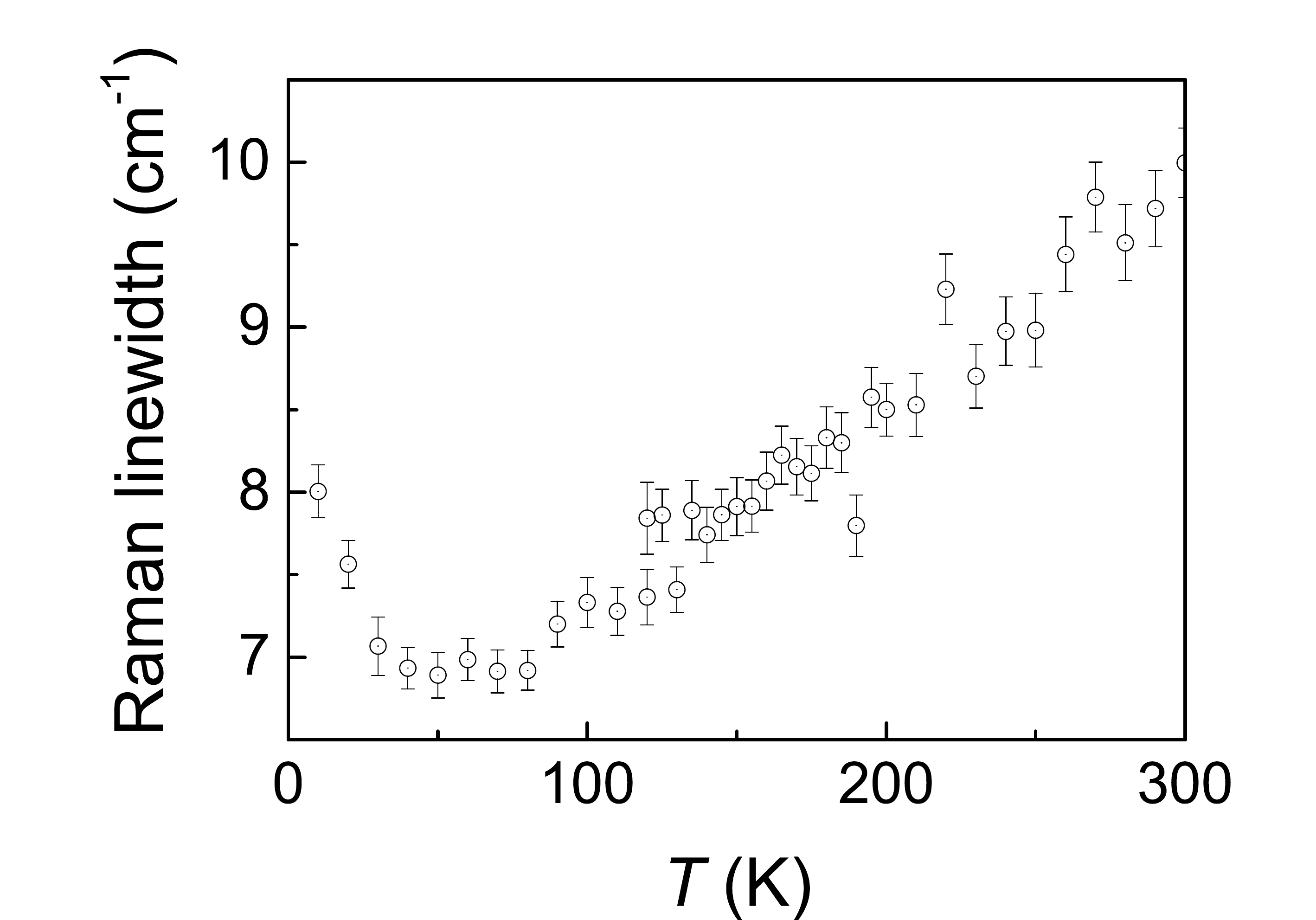}
	\caption{(color online) Common Lorentzian linewidth (FWHM) of the $A_g$ - $B_{1g}$ doublet observed near 380 cm$^{-1}$.}
	\label{RamanFWHM}
\end{figure}

Fig. \ref{Raman1} displays the Raman shift of the $A_g$ - $B_{1g}$ doublet near 380 cm$^{-1}$.
In contrast to the $\sim$160 cm$^{-1}$ $B_{2g}$ mode, magnetic effects in Raman modes at higher frequencies are either weaker or cannot be verified due to an increased linewidth of theses modes. Fig. \ref{RamanFWHM} shows the Lorentzian FWHM of these two modes. However, the Raman shifts as well as the linewidth show a maximum and a minimum centered at  $\sim$ 75 K similar to the behavior of the 160 cm$^{-1}$ $B_{2g}$ mode.

Similarly, the increase/decrease of the shift/linewidth towards very low temperatures correspond to the negative thermal expansion indicated by the temperature dependence of the lattice parameter $b$  and the anomalous length change at these temperatures. At these temperatures, magnetic effects of the Cr ordering can safely be ruled out since the Cr moments have reached saturation at these temperatures (see Figure \ref{N5PowLaw}).
To what extend, the Eu$^{3+}$ magnetism and especially the temperature dependent of thermal population of the $^7$F$_J$ ($J \neq$ 0) levels may be responsible for the observed effect remains elusive at present.

\subsection{Dielectric Properties}
\begin{figure}[htp]
	\centering
	\includegraphics[width=8.0cm]{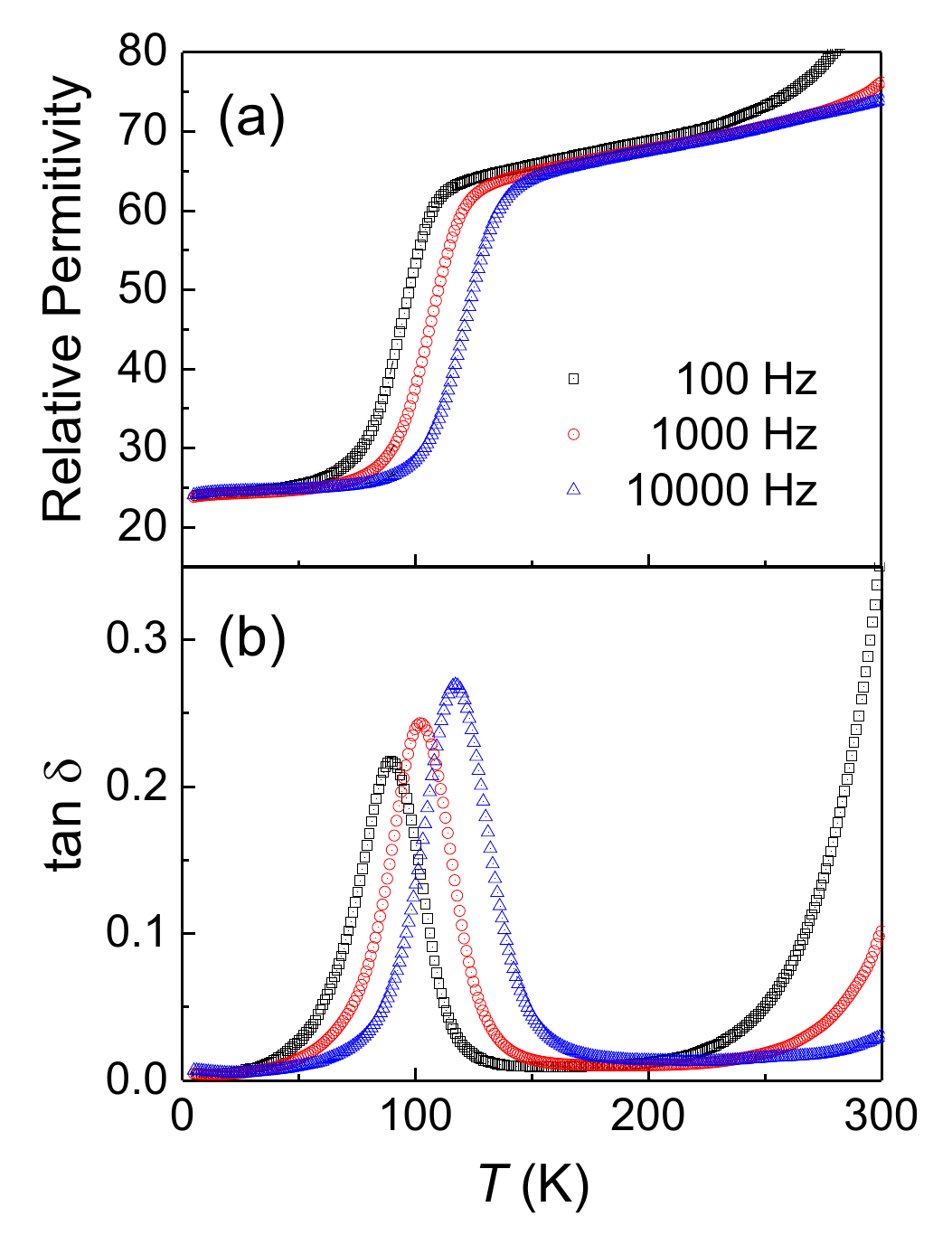}
	\caption{(color online) Temperature dependence of (a) the relative permittivity and (b) tan ($\delta $) (loses) of  polycrystalline sample of EuCrO$_3$ for several frequencies.}
	\label{Dielectric}
\end{figure}

The structural anomaly detected by the thermal expansion at $T \sim$ 100 K has a corresponding anomaly in the dielectric properties. Fig. \ref{Dielectric}(a) and (b) show the relative permittivity and the dielectric loss $tan(\delta)$ of EuCrO$_3$, measured at frequencies of 100, 1000 and 10000 Hz. On cooling, the relative permittivity (Fig. \ref{Dielectric}(a)) exhibits a step-like feature which for 100 Hz starts at  $\sim$ 100 K and leads to a decrease of the permittivity by a factor of 2.5. By increasing the frequency, the step-like feature moves to higher temperatures. The permittivity step and its frequency dependence are accompanied by a broad resonance-like feature in dielectric loss $tan(\delta)$  in Fig. \ref{Dielectric}(b). The peak temperature of the  frequency dependent dielectric loss $tan(\delta)$ maximum at $T_{\rm max}$ range between $\sim$80 and 125 K. This temperature interval coincides with the position of the highest slope in the capacitance and the observed anomaly in thermal expansion Fig. \ref{ThExpa}.
Measurements in magnetic fields up to 5.5 T did not alter the observed anomalies in dielectric properties.

\begin{figure}[htp]
	\centering
	\includegraphics[width=8.0cm]{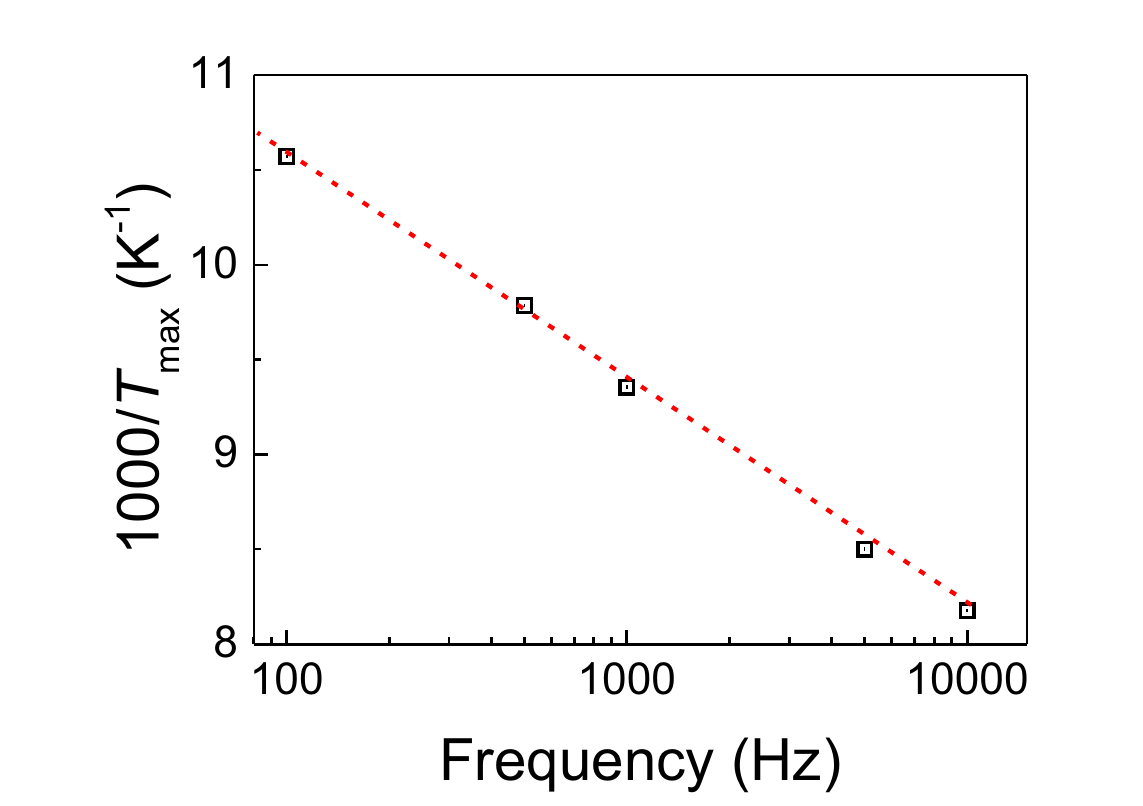}
	\caption{(color online) Frequency shift of the inverse maximum temperature in the conductance determined on a polycrystalline sample of EuCrO$_3$. The (red) dashed line represents an Arrhenius law according to Eq. (\ref{Tmax}) with an activation energy of 170 meV.}
	\label{TmaxActivated}
\end{figure}

The frequency shift of $T_{\rm max}$  exhibits an activated behavior (Fig. \ref{TmaxActivated}) according to
\begin{equation}
\nu(T) = \nu_0 e^{ (\frac{-\Delta E}{K_BT})},
\label{Tmax}
\end{equation}

with an activation energy, $\Delta E$, of

\begin{equation*}
\Delta E = 170(10) {\rm meV}
\end{equation*}

and an attempt frequency $\nu_0$ of

\begin{equation*}
\nu_0 \approx 1.3(8) \times 10^{11} {\rm Hz}.
\end{equation*}

\section {Discussion and Conclusion}
EuCrO$_3$ which crystallizes with the GdFeO$_3$ structure-type orders antiferromagnetically  below $\sim$ 178 K as we could show using a combination of magnetic susceptibility, heat capacity and NPD measurements. The magnetic susceptibility data reveal weak ferromagnetism  below the N\'{e}el temperature, in agreement with earlier studies.\cite{Tsushima1969,Gonjal2013} The magnetic contribution to the heat capacity is characterized by a $\lambda$-type anomaly at the N\'{e}el temperature containing a magnetic entropy in agreement with the entropy expected for a spin $S$ = 3/2 magnetic system consistent with the electronic configuration of the Cr$^{3+}$ cations.
Our XRPD and NPD measurements provide precise atomic positions of all atoms in the cell, especially for the light oxygen atoms.
The low-temperature NPD data revealed extra Bragg peaks of magnetic origin which can be attributed to a $G_x$ antiferromagnetic structure with an ordered moment of $\sim$ 2.4 $\mu_{\rm B}$ consistent with an $S$ =3/2 spin-only groundstate.
The weak ferromagnetic moment points along the $c$-axis and it amounts to $\sim$ 0.1, consistent with magnetization data. Evidence for an ordering of the Eu moments is neither found in the NPD data nor in the heat capacity data or magnetization data.

The contribution of the Eu$^{3+}$ cations to the magnetism  is rather characterized by a weakly temperature dependent Van Vleck-type paramagnetism arising from the  successive population of $J \neq$ 0 excited magnetic states.  Indication of a polarization of the Eu moments by the ordered Cr moments can be derived by the fits of the high-temperature ($T > T_{\rm N}$) magnetic susceptibility data.
Effects of the Eu magnetism could be the origin of the slight decrease of the saturation magnetization below 20 K (see inset Fig. \ref{Fig3}) and the decrease of the Raman lineshifts and the linewidth (see Figs. \ref{Raman160}, \ref{Raman1} and \ref{RamanFWHM}) and also the slight negative thermal expansion along the $b$ axis at low temperatures. (see Fig. \ref{XRDThExpa}).

Structural anomalies detected in the dielectric  Raman scattering and thermal expansion data indicate magnetoelastic distortion. These are clearly visible in the shift of some Raman modes at higher temperatures and also the dilatometer measurements of the lenght change carried out on a polycrystalline sample. Most striking is the anomalous contraction  of the $b$ lattice parameter below $T_{\rm N}$. It can be understood on the basis of the $G_x$-type antiferromagnetic structure with essentially antiparallel arrangement of the nearest neighbor Cr moments in the $a$ - $b$ plane and the tendency to strengthen the antiferromagnetic spin exchange coupling by decreasing the Cr - Cr distance. The decrease of the $b$ lattice parameter leads to a shortening of Cr - O2 distances, i.e. of the Cr to the oxygen atoms connecting the CrO$_6$ octahedra in the $a$ - $b$ plane leading also to a decrease of the nearest neighbor Cr - Cr distance in the $a$ - $b$ plane (3.84 \AA). Such a distance decrease leads to a strengthening of the superexchange interaction.\cite{deJongh1975,Shrivastava1976,Jansen1977}
The effect of the bond-length decrease is partially compensated by a  Cr - O2 - Cr bonding angle reduction. When decreasing $b$, the bonding angle is slightly moving  away from the favoring 180$^{\rm o}$ Cr - O - Cr configuration. This effect is obviously less determining than that induced by the distance decrease.
Along the $c$-axis, the decrease of the $b$ parameter also leads to a slight decrease of the Cr - O1 bond distance, i.e. of the Cr atoms to the apical oxygen atoms in the CrO$_6$ octahedra. The bonding angle Cr - O1 - Cr also slightly increases when $b$ shrinks.
Both effects favor an growth of the antiferromagnetic spin exchange along $c$. In total, the anomalous $b$ axis shortening starting below $T_{\rm N}$ is a consequence of the $G_x$-type antiferromagnetic ordering and the tendency to minimize the exchange energy and to strengthen the antiferromagnetic structure.

We finally discuss the anomalies seen in the dielectric permittivity and the conductance observed at temperatures around $\sim$100 K. The anomalies show activated behavior with an activation energy of 170 meV similar to the activation energy of a relaxor-like ferroelectric features detected in CeCrO$_3$, however at significantly higher temperatures.\cite{Shukla2009} For CeCrO$_3$ an activation energy of 130 meV was derived. Such energies correspond to typical optical phonon frequencies of oxygen related vibrations suggesting a vibrational origin of the dielectric anomalies. However, the calculated frequencies are largely different from Raman oxygen frequencies possibly indicating a different origin of the observed dielectric anomalies in both compounds.
The dielectric anomalies in EuCrO$_3$ are paralleled by anomalies in the thermal expansion indicating some origin in lattice anomalies which however could not be clearly detected by the less sensitive XRPD measurements. The absence of any magnetic field dependence also supports a lattice origin of the dielectric anomalies.

In summary, we have conclusively established the Cr antiferromagnetic ordering  and investigated and discussed associated anomalies in various lattice, thermal and electric properties in EuCrO$_3$. Our investigations add to complete the knowledge of the unusual physical and especially magnetic  properties of the rare earth orthochromite series.

\section*{Acknowledgment}

This work was financially supported by Brock University, the Natural Sciences and Engineering Research Council of Canada (NSERC), the Ministry of Research and Innovation (Ontario), and the Canada Foundation for Innovation, Canada. We thank E. Br\"ucher and G. Siegle for  expert experimental assistance and J. Rodr\'{\i}guez-Carvajal for advice with the size/shape refinement of the the particle size.

\end{document}